\documentclass[article]{JHEP3}

\usepackage{amssymb}
\usepackage{amsmath}
\usepackage{amsfonts}
\usepackage{graphicx}
\usepackage{epsfig}

\providecommand{\U}[1]{\protect\rule{.1in}{.1in}}

\newcommand{\be}{\begin{equation}}
\newcommand{\ee}{\end{equation}}
\newcommand{\bea}{\begin{eqnarray}}
\newcommand{\eea}{\end{eqnarray}}
\newcommand{\bean}{\begin{eqnarray*}}
\newcommand{\eean}{\end{eqnarray*}}

\def\beq{\begin{equation}}

\def\eeq{\end{equation}}

\relax

\preprint{ {\small{\textsf{ROM2F/2006/25}}}
\\
{\small{\textsf{LPTENS-06/49}}}
\\
{\small{\textsf{CERN-PH-TH/2006-233}}}
}

\title{Eikonal Approximation in AdS/CFT:
\\
From Shock Waves to Four--Point Functions}

\author{Lorenzo Cornalba$^a$, Miguel S. Costa$^{b,c}$, Jo\~ao Penedones$^{b,c}$ and Ricardo Schiappa$^d$
\\
\\
$^{a}$Dipartimento di Fisica \& INFN, Universit\'{a} di Roma ``Tor Vergata'',\\
Via della Ricerca Scientifica 1, 00133, Roma, Italy\\
\\
$^{b}$Departamento de F\'{i}sica e Centro de F\'{i}sica do Porto,\\
Faculdade de Ci\^{e}ncias da Universidade do Porto,\\
Rua do Campo Alegre, 687, 4169--007 Porto, Portugal\\
\\
$^{c}$Laboratoire de Physique Th\'eorique de l'Ecole Normale
Sup\'erieure\footnote{Unit\'e mixte du C.N.R.S. et de l'Ecole Normale Sup\'erieure, UMR 8549.},\\
24 Rue Lhomond, 75231 Paris, France\\
\\
$^{d}$Theory Division, Department of Physics, CERN,\\
CH--1211 Gen\`eve 23, Switzerland\\
\\
\email{cornalba@roma2.infn.it},\quad
\email{miguelc@fc.up.pt},\quad
\email{jpenedones@fc.up.pt},\quad
\email{ricardos@mail.cern.ch}
}

\abstract{We initiate a program to generalize the standard eikonal approximation to compute amplitudes in
Anti--de Sitter spacetimes. Inspired by the shock wave derivation of the eikonal amplitude in flat space,
we study the two--point function $\mathcal{E} \sim \left\langle \mathcal{O}_{1} \mathcal{O}_{1} \right\rangle_{\mathrm{shock}}$
in the presence of a shock wave in Anti--de Sitter, where $\mathcal{O}_{1}$ is a scalar primary operator in the dual
conformal field theory. At tree level in the gravitational coupling, we relate the shock two--point
function $\mathcal{E}$ to the discontinuity across a kinematical branch cut of the conformal field theory four--point
function $\mathcal{A} \sim \left\langle \mathcal{O}_{1} \mathcal{O}_{2} \mathcal{O}_{1} \mathcal{O}_{2} \right\rangle$,
where $\mathcal{O}_2$ creates the shock geometry in Anti--de Sitter. Finally, we extend the above results by computing
$\mathcal{E}$ in the presence of shock waves along the horizon of Schwarzschild BTZ black holes. This work gives new tools
for the study of Planckian physics in Anti--de Sitter spacetimes.}


\begin{document}



\vfill

\eject


\section{Introduction and Summary}


The $\mathrm{AdS}_{d+1}$/$\mathrm{CFT}_{d}$ correspondence relates, in general, a theory of strings on the negatively curved Anti--de Sitter (AdS) space with a conformal field theory (CFT) living on its boundary \cite{Malda1, WittenGubser, w98, Malda2}. When the radius $\ell$ of $\mathrm{AdS}_{d+1}$ is large compared with the string length $\ell_{s}$, we can, in first approximation, analyze the dynamics of the low--energy gravitational theory for the massless string modes. However, in most circumstances, we are forced to restrict our attention to tree level gravitational interactions, since the loop expansion in the gravitational coupling $G$ is plagued with the usual ultra--violet (UV) problems present also in flat space, when we neglect the regulator length $\ell_{s}$. In the prototypical example of the duality between type IIB strings on $\mathrm{AdS}_{5} \times {S}_{5}$ and $\mathcal{N}=4$ $U\left( N \right)$ supersymmetric Yang--Mills (SYM) theory in $d=4$, the gravitational coupling in units of the $\mathrm{AdS}$ radius $G\ell^{-3}$ is proportional to $N^{-2}$, and therefore we are in general forced to consider the planar limit of the SYM theory, even when the 't Hooft coupling $\left( \ell/\ell_{s}\right)^{4}$ is large. Moreover, even at tree level, Feynman graphs which are readily computed in flat space are extremely complex in $\mathrm{AdS}$, limiting the practical use of the perturbative expansion \cite{FreedmanRev, Rastelli1, dmmr99}.

In this paper we initiate a program to go beyond the tree level approximation and to explore the physics on $\mathrm{AdS}_{d+1}$ at \textit{finite} $G\ell^{1-d}$. To do so, we recall that, in flat space $\mathbb{M}^{d+1}$, the quantum effects of various types of interactions can be reliably re--summed to all orders in the relevant coupling constant, in specific kinematical regimes \cite{LevySucher, tHooft, Kabat, ACV, tHooft3d, DeserJackiw, Deser}. In particular, the amplitude for the scattering of two particles can be approximately computed in the eikonal limit of small momentum transfer compared to the center--of--mass energy, or, equivalently, of small scattering angle. In this limit, even the gravitational interaction can be approximately evaluated to \textit{all} orders in $G$, and the usual perturbative UV problems are rendered harmless by the re--summation process. Moreover, at large energies, the gravitational interaction dominates all other interactions, quite independently of the underlying theory \cite{tHooft}. At high--energies, scattering amplitudes in the eikonal limit exhibit a universal behavior which is indicative of the presence of gravity in the theory under consideration.

It is therefore tempting to speculate that, in certain favored kinematical regimes, quantum effects can be re--summed also in $\mathrm{AdS}$ and that the gravitational interaction, if present, will dominate all other interactions and exhibit a universal behavior which will be a clear signal
of the existence of a gravitational description in the dual $\mathrm{CFT}$. This paper is a first step towards the consistent application of eikonal methods to the dynamics in $\mathrm{AdS}$ and to the physics of the dual $\mathrm{CFT}$.

Let us start by recalling some basic facts about the eikonal formalism in flat space. Consider the scattering of two scalar particles in flat Minkowski space $\mathbb{M}^{d+1}$. For the present purposes, we work at high--energies and we neglect the masses of the scattering particles. The scattering amplitude $\mathcal{A}$ is a function of the Mandelstam invariants $s$ and $t$ and is computed in perturbation theory
\begin{figure}
[ptb]
\begin{center}
\includegraphics[width=3.8685in]
{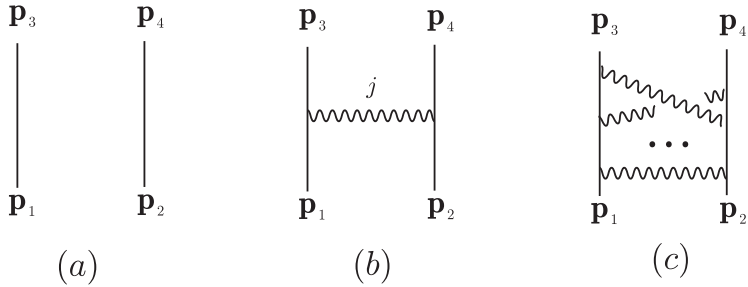}
\caption{Interaction diagrams in both flat and AdS spaces. In the eikonal regime, free propagation $(a)$ is modified primarily by interactions described by crossed--ladder graphs $(c)$. In flat space and in this regime, the tree level amplitude is dominated by the \emph{T}--channel graph $(b)$ with maximal spin $j=2$ of the exchanged massless particle. Moreover, the full eikonal amplitude can be computed from diagram $(b)$.}
\label{fig4}
\end{center}
\end{figure}
\[
\mathcal{A} = \mathcal{A}_{0}+\mathcal{A}_{1}+\cdots~,
\]
where $\mathcal{A}_{0}$ corresponds to graph $\left( a\right)$ in figure \ref{fig4} describing free propagation in spacetime. The tree level amplitude $\mathcal{A}_{1}$ contains, in general, many different graphs. However, in the eikonal regime of small scattering angle $-t\ll s$, the \emph{T}--channel exchange of massless particles dominates the full tree level amplitude. Therefore, the only relevant contribution to $\mathcal{A}_{1}$ will come from graph $(b)$ in figure \ref{fig4}, where $j$ denotes the spin of the exchanged massless particle. More precisely, this contribution reads \footnote{The normalization $4^{3-j}~2\pi i\,G$ has been chosen for later convenience. When $j=2$, the gravitational coupling constant $G$ is the canonically normalized Newton constant.}
\[
\mathcal{A}_{1} \simeq 4^{3-j}~2\pi i\,G~\frac{s^{j}+c_{1}s^{j-1}t+\cdots +c_{j}t^{j}}{-t}\ .
\]
In the eikonal limit, the full amplitude $\mathcal{A}$ is dominated by the ladder graphs $(c)$ in figure \ref{fig4} and can be reconstructed starting from $\mathcal{A}_{1}$. More precisely, we write $\mathcal{A}$ in the impact--parameter representation
\begin{equation}
\mathcal{A}\left(  s,t=-q^{2}\right)  \,\simeq\,2s\int_{\mathbb{E}^{d-1}}dx\;e^{iq\cdot x}\;e^{-2\pi i\,\sigma(s,r)}\ ,\label{eikonalresum}
\end{equation}
where $r=\sqrt{x^{2}}$ is the radial coordinate in transverse space $\mathbb{E}^{d-1}$ and $q$ is the transverse momentum transfer. In general, the phase shift $\sigma\left( s,r\right)$ receives contributions at all orders in perturbation theory. However, the leading behavior of $\sigma\left( s,r\right)$ for large $r$ is uniquely determined by the tree level interaction $\mathcal{A}_{1}$ and is therefore obtained by a simple Fourier transform
\begin{equation}
\mathcal{A}_{1}\left(  s,t=-q^{2}\right)  \,\simeq\,-4\pi is\int_{\mathbb{E}^{d-1}}dx\;e^{iq\cdot x}\;\sigma(s,r)\ .\label{imprep1}
\end{equation}
This yields
\[
\sigma\left(  s,r\right)  \simeq-8G\left(  \frac{s}{4}\right)  ^{j-1}
\Pi\left(  r\right)  ~,
\]
where $\Pi\left( r\right)$ is the massless Euclidean propagator in transverse space $\mathbb{E}^{d-1}$. The behavior of $\sigma$ for large $r$ is determined only by the residue of the $1/t$ pole in $\mathcal{A}_{1}$, and is it insensitive to the other terms, proportional to $c_{i}$, which are regular for $t\rightarrow0$. Hence, the higher order graphs of figure \ref{fig4}$(c)$ are taken into account simply by exponentiating the phase $\sigma$ in (\ref{eikonalresum}). In the limit of high--energy $s$, the mediating massless particle with maximal spin $j$ dominates the interaction. In theories of gravity, this particle is the graviton, with $j=2$.

In the  literature there are essentially two derivations of the eikonal amplitude (\ref{eikonalresum}). One derivation \cite{LevySucher} considers the behavior of the Feynman diagrams in figure \ref{fig4}$(c)$ in the limit $-t\ll s$, which, after a careful combinatorial analysis, re--sum to the result (\ref{eikonalresum}). The second derivation \cite{tHooft} is more geometrical and considers the motion of particle 1 in the classical field configuration created by particle 2. In the limit of large $s$, the particles move approximately at the speed of light, and particle 2 is viewed as a source, localized along its null world--line, for the exchanged massless spin $j$ field. This classical source produces a shock wave configuration \cite{Aichelburg:1970dh} in the exchanged field, and one may solve the wave equation for particle 1 in the presence of this classical background. When crossing the shock wave, the phase of the wave function for particle 1 is shifted by $-2\pi\sigma(r
 ,s)$ and the amplitude between the initial and final states of particle 1 is then given by the eikonal result (\ref{eikonalresum}).

In this work, we shall consider the $\mathrm{CFT}$ analogue of $2\rightarrow2$ scattering of scalar fields in flat space. More precisely, we will study the $\mathrm{CFT}$ correlator
\[
\left\langle \mathcal{O}_{1}\left( \mathbf{p}_{1}\right) \mathcal{O}_{2}\left( \mathbf{p}_{2}\right) \mathcal{O}_{1}\left( \mathbf{p}_{3}\right) \mathcal{O}_{2}\left( \mathbf{p}_{4}\right) \right\rangle_{\text{\textrm{CFT}}_{d}} \equiv \frac{1}{\mathbf{p}_{13}^{\Delta_{1}}\mathbf{p}_{24}^{\Delta_{2}}}\,\mathcal{A}\left( z,\bar{z}\right)~,
\]
where the scalar primary operators $\mathcal{O}_{1}$, $\mathcal{O}_{2}$ have conformal dimensions $\Delta_{1}$, $\Delta_{2}$, respectively. The $\mathbf{p}_{i}$ are now points on the boundary of $\mathrm{AdS}$, and the amplitude $\mathcal{A}$ is a function of two cross--ratios $z$,$\bar{z}$ (the precise definition of our notation can be found in section \ref{notation}). Neglecting string effects, the function $\mathcal{A}$ can be computed in the dual $\mathrm{AdS}$ formulation as a field theoretic perturbation series \cite{WittenGubser, w98, Rastelli1, dmmr99}
\[
\mathcal{A}=\mathcal{A}_{0}+\mathcal{A}_{1}+\cdots~,
\]
where $\mathcal{A}_{0}=1$ corresponds to free propagation described by the Witten diagram in figure \ref{fig4}$(a)$. We expect that the eikonal kinematical regime in AdS is still defined by $\mathbf{p}_{1}\sim\mathbf{p}_{3}$, which corresponds to the limit of small cross--ratios $z$,$\bar{z}$. In analogy with flat space, we shall focus uniquely on the contribution to $\mathcal{A}_{1}$ coming from the graph \ref{fig4}$(b)$.

\begin{figure}
[ptb]
\begin{center}
\includegraphics[height=1.2256in, width=2.176in]
{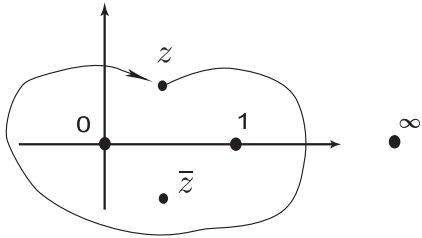}
\caption{Analytic continuation of $\mathcal{A}_{1}$ to obtain $\mathcal{A}_{1}^{\circlearrowright}$. The variable $\bar{z}$ is kept fixed and $z$ is transported clockwise around the point at infinity, circling the points $0,1,\bar{z}$.}
\label{fig9}
\end{center}
\end{figure}

The direct generalization of the flat space eikonal re--summation to AdS is not obvious, because AdS graphs are much harder to compute even at tree level \cite{Rastelli1, dmmr99}. Fortunately, as described above, in flat space there is an alternative way to derive the eikonal result (\ref{eikonalresum}), which uses the shock wave geometry of the exchanged massless field. In this paper, we shall extend this analysis by considering the two--point function $\mathcal{E} \sim \left\langle \mathcal{O}_{1}\mathcal{O}_{1}\right\rangle _{\mathrm{shock}}$ on $\mathrm{AdS}_{d+1}$ in the presence of a shock wave of a spin $j$ massless field. By analogy with flat space, we expect that the shock wave two--point function $\mathcal{E}$ contains contributions from all ladder graphs of figure \ref{fig4}. In particular, the first two terms in the coupling constant expansion
\[
\mathcal{E}=\mathcal{E}_{0}+\mathcal{E}_{1}+\cdots~,
\]
should correspond, respectively, to free propagation and to tree level \emph{T}--channel exchange of a spin $j$ massless particle. Indeed, we shall determine a precise relation between $\mathcal{E}_{1}$ and the tree level amplitude $\mathcal{A}_{1}$ associated to graph \ref{fig4}$(b)$. We will find that $\mathcal{E}_{1}$ controls the small $z$,$\bar{z}$ behavior of the discontinuity function (monodromy)
\begin{equation}
\mathcal{M}_{1} \left( z,\bar{z}\right) =
\operatorname{Disc}{}_{z} ~\mathcal{A}_{1} \left( z,\bar{z}\right)
\equiv \frac{1}{2\pi i} \Big( \mathcal{A}_{1}^{\circlearrowright}
\left( z,\bar{z}\right) - \mathcal{A}_{1} \left( z,\bar{z}\right)
\Big)\ , \label{xxxyyy}
\end{equation}
where $\mathcal{A}_{1}^{\circlearrowright}$ is the analytic continuation of $\mathcal{A}_{1}$ obtained by keeping $\bar{z}$ fixed and by transporting $z$ clockwise around the point at infinity as in figure \ref{fig9}. More precisely, the small $z$,$\bar{z}$ behavior of $\mathcal{M}_{1}$ plays the role of the residue of the $1/t$ pole in $\mathcal{A}_{1}$ in the flat space case and it is given by
\begin{equation}
\mathcal{M}_{1}\left(  z,\bar{z}\right)  \simeq\left(  -z\right)
^{1-j}M\left(  \frac{\bar{z}}{z}\right)  ~,\label{notOPE}
\end{equation}
where the function $M(w)$ satisfies
\begin{align*}
M\left( w\right) & = w^{1-j}M\left( 1/w\right) ~,\\
M\left( w\right)^{\star} & = M\left( w^{\star}\right) ~.
\end{align*}
Note that (\ref{notOPE}) is not directly related to the small $z$,$\bar{z}$ behavior of $\mathcal{A}_{1}$, which is in turn controlled by the standard OPE in the dual CFT. As in flat space, the maximal spin $j$ dominates in the eikonal regime of small $z,\bar{z}$. The leading behavior of $\mathcal{M}_{1}$ is explicitly given as an integral representation over transverse space analogous to (\ref{imprep1}). As derived in section \ref{secsix}, we shall find that
\begin{align}
\mathcal{M}_{1} & \simeq -8G \mathcal{N}_{\Delta_{1}} \mathcal{N}_{\Delta_{2}} \Gamma \left( 2\Delta_{1}-1+j \right) \Gamma \left( 2\Delta_{2}-1+j \right) \times \label{tt11}\\
& \times~\left( -q^{2}\right)^{\Delta_{1}} \left( -p^{2}\right)^{\Delta_{2}} \int_{H_{d-1}} \widetilde{dx} \widetilde{dy}~\frac{\Pi\left( x,y \right)}{\left( 2q\cdot x \right)^{2\Delta_{1}-1+j} \left( 2p\cdot y \right)^{2\Delta_{2}-1+j}}~,\nonumber
\end{align}
where the relevant notation is given in detail in section \ref{notation}. In a companion paper \cite{Paper2} we will explore the CFT consequences of the above result.

One crucial step is missing to complete the eikonal program. In flat space, one can approximately reconstruct the full amplitude from the tree level phase shift $\sigma$ using (\ref{eikonalresum}). This last step cannot be immediately done in $\mathrm{AdS}_{d+1}$, since the tree level eikonal two--point function $\mathcal{E}_{1}$ is not related to $\mathcal{A}_{1}$ but rather to the \textit{discontinuity} $\mathcal{M}_{1}$ of $\mathcal{A}_{1}$ across a kinematical branch cut. Therefore, in order to reconstruct $\mathcal{A}_{1}$ and the full amplitude $\mathcal{A}$ from the shock wave two--point function, extra information is needed. In \cite{Paper2} we shall conjecture a possible resolution of this problem, even though more work is needed to put these results on firm grounds.

Let us further comment on the above result, and on its relation to the more familiar flat space case. In the CFT literature, one usually considers the amplitude $\mathcal{A}_{1}\left( z,\bar{z}\right)$ evaluated on the principal Euclidean sheet, where $\bar{z}=z^{\star}$ (again see section \ref{notation} for a precise definition of the notation). It is quite clear, on the other hand, that eikonal results, both in flat and in AdS space, probe amplitudes deep in the Lorentzian regime. In fact, scattered particles are almost light--like separated in position space, due to the high relative energies. Therefore, in the context of the AdS/CFT duality, we must also consider the boundary CFT correlator $\mathcal{A}_{1}\left( z,\bar{z}\right)$ in its Lorentzian regime. This is obtained by analytically continuing $\mathcal{A}_{1}\left( z,\bar{z}\right)$, with $z,\bar{z}$ now viewed as independent variables. The channel $z,\bar{z}\rightarrow 0$ corresponds to light--like separation of the relevant scattering particles but, as explained in detail in section \ref{CreateShock}, this limit must be accompanied by the relevant analytic continuation in (\ref{xxxyyy}). Note that, implicitly, this continuation is also performed in the flat space eikonal computation, but it is immaterial in this case since, in momentum space, tree level amplitudes are rational functions of the kinematical invariants. Finally, let us note that, without analytic continuation, the limit of $\mathcal{A}_{1}$ for $z,\bar{z}\rightarrow 0$ goes like $\left\vert z\right\vert ^{\mathrm{Energy}}$ and is therefore governed by the usual OPE. In this case, the relevant contribution comes from only a finite number of states of lowest energy propagating in this channel . This corresponds to the state $\mathcal{O}_{j}$ dual to the the spin $j$ particle being exchanged, and the eikonal result would amount to a computation of the three--point coupling $\left\langle \mathcal{O}_{1}\mathcal{O}_{1}\mathcal{O}_{j}\right\rangle \left\langle \mathcal{O}_{j} \mathcal{O}_{2}\mathcal{O}_{2}\right\rangle$. The eikonal result, on the other hand, contains much more information. The limit $z,\bar{z}\rightarrow 0$ of $\mathcal{M}_{1}$ goes in fact like $\left\vert z\right\vert ^{1-\mathrm{Spin}}$ and we obtain information about the \textit{full tower of spin }$j$ \textit{intermediate states}, which is encoded in the generating function $M\left( w\right)$.

Finally, in section \ref{BTZsec}, we extend the computation of the two--point function $\mathcal{E} \sim \left\langle \mathcal{O}_{1}\mathcal{O}_{1} \right\rangle_{\mathrm{shock}}$ to the case of a shock wave propagating along the horizon of a Schwarzschild BTZ black hole \cite{BTZ, BTZ22}. This computation extends the results of \cite{kos02, fhks03, fl05, br03, hlr06}, where CFT correlators are used to extract information on the physics behind the horizon of the black hole, with particular emphasis on the singularity. In particular, in future work we plan to relate $\mathcal{E}$ to the four--point function in the BTZ geometry at all orders in $G$, thus probing the physics of the singularity in a truly quantum gravity regime.

In two appendices, we further include a full discussion on the AdS and Hyperbolic space propagators required in the main text; and an explicit calculation of the shock two--point function, in the case of $d=3$, using Poincar\'e coordinates. This will provide the reader who is familiar with the correlation function calculations of \cite{w98, fmmr98} a simpler access to the calculations we perform in the bulk of the paper.


\section{Preliminaries and Notation\label{notation}}

\begin{figure}
[ptb]
\begin{center}
\includegraphics[width=2.176in]
{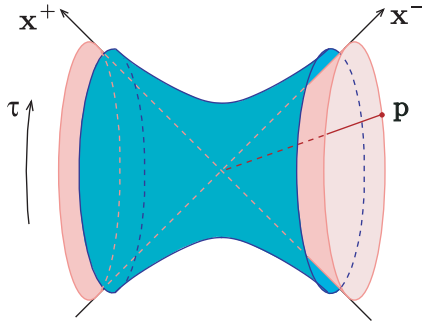}
\caption{Embedding of AdS$_2$ in $\mathbb{M}^{2}\times\mathbb{M}^{1}$. A point  $\mathbf{p}$ in the boundary of AdS$_2$ is a null ray in $\mathbb{M}^{2}\times\mathbb{M}^{1}$.}
\label{AdSTwo}
\end{center}
\end{figure}

Recall that \textrm{AdS}$_{d+1}$ space, of dimension $d+1$, can be defined as a pseudo--sphere in the embedding space $\mathbb{M}^{2} \times \mathbb{M}^{d}$. We denote with $\mathbf{x}=\left(  x^{+},x^{-},x\right)$ a point in $\mathbb{M}^{2} \times \mathbb{M}^{d}$, where $x^{\pm}$ are light cone coordinates on $\mathbb{M}^{2}$ and $x$ denotes a point in $\mathbb{M}^{d}$. Then, AdS space of radius $\ell$ is described by \footnote{We denote with $\mathbf{x}\cdot\mathbf{y}$ and $x\cdot y$ the scalar products in $\mathbb{M}^{2}\times\mathbb{M}^{d}$ and $\mathbb{M}^{d}$, respectively. Moreover, we abbreviate $\mathbf{x}^{2}\mathbf{=x}\cdot \mathbf{x}$ and $x^{2}=x\cdot x$ when clear from context. In $\mathbb{M}^{d}$ we shall use coordinates $x^\mu$ with $\mu=0,\ldots,d-1$ and with $x^0$ the timelike coordinate. Finally, in $\mathbb{M}^{2}$ we shall write $x^{\pm}=x^{d+1}\pm x^{d}$ for the light cone coordinates, with $x^{d+1}$ the time direction and $x^{d}$ the spatial one.}
\begin{equation}
\mathbf{x}^{2}=-x^{+}x^{-}+x^{2}=-\ell^{2}~.\label{pseudoS}
\end{equation}
Similarly, a point on the holographic boundary of \textrm{AdS}$_{d+1}$ can be described by a ray on the light cone in $\mathbb{M}^{2}\times\mathbb{M}^{d}$, that is by a point $\mathbf{p}$ with
\[
\mathbf{p}^{2}=-p^{+}p^{-}+p^{2}=0~,
\]
defined up to re--scaling
\[
\mathbf{p}\sim\lambda \mathbf{p} \qquad \left( \lambda>0\right).
\]
In figure \ref{AdSTwo} the embedding of the AdS geometry is represented for the AdS$_2$ case. From now on we choose units such that $\ell=1$.

The AdS/CFT correspondence predicts the existence of a dual CFT$_{d}$ living on the boundary of AdS$_{d+1}$. In particular, a CFT correlator of scalar primary operators located at points $\mathbf{p}_{1},\ldots,\mathbf{p}_{n}$, can be conveniently described by an amplitude
\[
A\left(  \mathbf{p}_{1},\ldots,\mathbf{p}_{n}\right)
\]
invariant under $SO\left(  2,d\right)  $ and therefore only a function of the invariants
\[
\mathbf{p}_{ij}=-2\mathbf{p}_{i}\cdot\mathbf{p}_{j}~.
\]
Since the boundary points $\mathbf{p}_{i}$ are defined only up to rescaling, the amplitude $A$ will be homogeneous in each entry
\[
A\left( \ldots, \lambda\mathbf{p}_{i}, \ldots \right) = \lambda^{-\Delta_{i}} A\left( \ldots, \mathbf{p}_{i}, \ldots \right)~,
\]
where $\Delta_{i}$ is the conformal dimension of the $i$--th scalar primary operator.

Throughout this paper we will focus our attention on four--point amplitudes of scalar primary operators. More precisely, we shall consider correlators of the form
\[
A\left( \mathbf{p}_{1},\mathbf{p}_{2},\mathbf{p}_{3},\mathbf{p}_{4} \right) = \left\langle \mathcal{O}_{1}\left( \mathbf{p}_{1}\right)  \mathcal{O}_{2}\left( \mathbf{p}_{2}\right) \mathcal{O}_{1}\left( \mathbf{p}_{3}\right) \mathcal{O}_{2}\left( \mathbf{p}_{4}\right) \right\rangle_{\text{\textrm{CFT}}_{d}}
\]
where the scalar operators $\mathcal{O}_{1},\mathcal{O}_{2}$ have dimensions
\[
\Delta_{1}=\Delta+\nu~, \qquad \Delta _{2}=\Delta-\nu~,
\]
respectively. The four--point amplitude $A$ is just a function of two cross--ratios $z,\bar{z}$ which we define, following \cite{Osborn, Osborn22}, in terms of the kinematical invariants $\mathbf{p}_{ij}$ as \footnote{Throughout the paper, we shall consider barred and unbarred variables as independent, with complex conjugation denoted by $\star$. In general $\bar{z}=z^{\star}$ when considering the analytic continuation of the CFT$_{d}$ to Euclidean signature. For Lorentzian signature, either $\bar{z}=z^{\star}$ or both $z$ and $\bar{z}$ are real. These facts follow simply from solving the quadratic equations for $z$ and $\bar{z}$.}
\begin{align*}
z\bar{z}  & =\frac{\mathbf{p}_{13}\mathbf{p}_{24}}{\mathbf{p}_{12}
\mathbf{p}_{34}}~,\\
\left(  1-z\right)  \left(  1-\bar{z}\right)   & =\frac{\mathbf{p}
_{14}\mathbf{p}_{23}}{\mathbf{p}_{12}\mathbf{p}_{34}}~.
\end{align*}
Then, the four--point amplitude can be written as
\[
A\left(  \mathbf{p}_{i}\right)  =\frac{1}{\mathbf{p}_{13}^{\Delta_{1}
}\mathbf{p}_{24}^{\Delta_{2}}}\mathcal{A}\left(  z,\bar{z}\right)~,
\]
where $\mathcal{A}$ is a generic function of $z,\bar{z}$. By conformal invariance, we can fix the position of up to three of the external points $\mathbf{p}_{i}$. In what follows, we shall often choose the external kinematics by placing the four points $\mathbf{p}_{i}$ at
\begin{align}
\mathbf{p}_{1}  & =\left(  0,1,0\right) ~,\ \ \ \ \ \ \ \ \ \ \ \
\ \ \ \ \ \ \ \ \ \ \ \mathbf{p}_{2}=-\left(
1,p^{2},p\right)  ~,\label{extpoints}\\
\mathbf{p}_{3}  & =-\left(  q^{2},1,q\right) ~,\ \ \ \ \ \ \ \ \ \
\ \ \ \ \ \ \ \ \ \mathbf{p}_{4}=\left(  1,0,0\right) ~,\nonumber
\end{align}
and we shall view the amplitude as a function of $p,q\in\mathbb{M}^{d}$. The cross ratios $z,\bar{z}$ are in particular determined by
\[
z\bar{z}=q^{2}p^{2}, \qquad
z+\bar{z}=2p\cdot q~.
\]
When $d=2$ it is convenient to parameterize $\mathbb{M}^{d}$ by light cone coordinates $x=x^0+x^1$ and $\bar{x}=x^0-x^1$, with metric $-dxd\bar{x}$. Then, if we choose $p=\bar{p}=-1$ we have $q=z$, $\bar{q}=\bar{z}$.

In the sequel, we will denote with $H_{d-1}\subset\mathbb{M}^{d}$ the transverse hyperbolic space, given by the upper mass--shell
\[
x^{2}=-1~ \qquad \left(  x^{0}>0\right),
\]
where $x\in\mathbb{M}^{d}$. We will also denote with $\mathrm{M}\subset\mathbb{M}^{d}$ the future Milne wedge given by $x^{2}\leq0$, \thinspace$x^{0}\geq0$. Similarly, we denote with $-\mathrm{M}$ the past Milne wedge and with $-H_{d-1}$ the associated transverse hyperbolic space. Finally we denote with
\begin{align*}
\widetilde{d\mathbf{x}} & =2d\mathbf{x}~\delta\left(  \mathbf{x}
^{2}+1\right)  ,\ \ \ \ \ \ \ \ \ \ \ \\
\widetilde{dx}  & =2dx~\delta\left(  x^{2}+1\right)  ,
\end{align*}
the volume elements on \textrm{AdS}$_{d+1}$ and $H_{d-1}$, respectively. For example
\[
\int_{\mathrm{M}}dx  =
\int_{0}^{\infty}t^{d-1}dt\int_{H_{d-1}
}\widetilde{dy}~,
\]
where $x=ty$ and $y \in H_{d-1}$.

Throughout the paper, we will often need the massless Minkowskian scalar propagator $\mathbf{\Pi}\left( \mathbf{x,y}\right)$ on \textrm{AdS}$_{d+1}$ and the massive Euclidean scalar propagator $\Pi\left( x,y\right)$ on $H_{d-1}$ of mass--squared $d-1\,$. They are canonically normalized by
\begin{align*}
\square_{\mathrm{AdS}_{d+1}}\,\mathbf{\Pi}\left(
\mathbf{x,y}\right)   &
=i~\mathbf{\delta}\left(  \mathbf{x,y}\right)  ~,\\
\left[  \square_{H_{d-1}}-\left(  d-1\right)  \right]
\,\Pi\left( x,y\right)   & =-~\delta\left(  x,y\right)  ~,
\end{align*}
and are explicitly given in appendix A. We also introduce, for future use, the constant $\mathcal{N}_{\Delta}$ given by the integral
$$
\mathcal{N}_{\Delta}^{-1} = \int_{\mathrm{M}} \frac{dy}{\left\vert y \right\vert
^{d-2\Delta}}~~e^{2k\cdot y} = \Gamma\left( 2\Delta\right) \int_{H_{d-1}} \frac{\widetilde{dy}}{\left( -2k\cdot y\right)^{2\Delta}} = \frac{\pi^{\frac{d}{2}-1}}{2} \Gamma\left( \Delta\right) \Gamma\left( \Delta-\frac{d}{2}+1\right)  ~.
$$

\begin{figure}
[ptb]
\begin{center}
\includegraphics[keepaspectratio,height=6cm]
{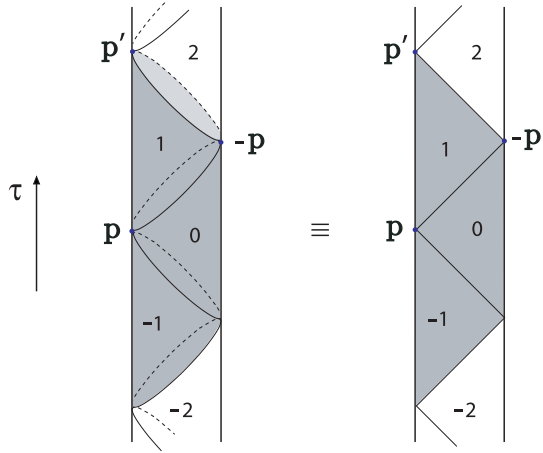}
\caption{Poincar\'e patches of an arbitrary boundary point $\mathbf{p}$, separated by the null surfaces $-2\mathbf{x\cdot p}=0$. Here AdS is represented as a cylinder with boundary $\mathbb{R}\times {S}_{d-1}$. Throughout this paper we shall mostly use a two--dimensional simplification of this picture, as shown in the figure. The point $-\mathbf{p}$ and an image point $\mathbf{p}'$ of $\mathbf{p}$ are also shown.}
\label{DecompAdS}
\end{center}
\end{figure}

To conclude this section, let us remind the reader that we have been careless about the global structure of \textrm{AdS}$_{d+1}$. As it is well known, the locus (\ref{pseudoS}) has a non--contractible timelike circle, and we shall denote with \textrm{AdS}$_{d+1}$ the covering space of (\ref{pseudoS}), where global time $\tau$, given by
\begin{equation}x^{d+1}+i x^0 =
\frac{1}{2}\left(  x^{+}+x^{-}\right)  +ix^{0}=\cosh\left(
\rho\right) ~e^{i\tau}~,\label{globalcoord}
\end{equation}
is decompactified. Therefore, one must be cautious when working in the embedding coordinates since two general bulk points $\mathbf{x}$ and $\mathbf{x'}$, or two boundary points $\mathbf{p}$ and $\mathbf{p'}$, related by a global time translation of integer multiples of $2\pi$ have the same embedding in $\mathbb{M}^{2}\times\mathbb{M}^{d}$. Given a boundary point $\mathbf{p}$, we may divide the \textrm{AdS} space in an infinite sequence of Poincar\'{e} patches separated by the null surfaces $\mathbf{x\cdot p}=0$ and labeled by integers $n$ increasing as we move forward in global time. The $n=0$ patch is the one which is spacelike related to the boundary point, as shown in figure \ref{DecompAdS}. These global issues will be relevant in sections 4 and 5.
%


\section{The Shock Wave Geometry\label{ShockSec}}


In this section we review the shock wave geometry in \textrm{AdS}$_{d+1}$, which is a direct analog of the Aichelburg--Sexl geometry in flat space and which has been described in \cite{dh85, ht93, pg97, Sfetsos, HI, Arcioni}.

\begin{figure}
[ptb]
\begin{center}
\includegraphics[height=1.966in]
{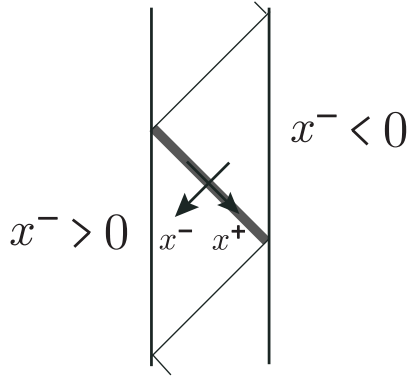}
\caption{Two consecutive Poincar\'{e} patches with $x^{-}>0$ and $x^{-}<0$. The shock geometry can be described by specifying gluing conditions on the separating surface $x^{-}=0$, which is parameterized by $x^{+}$ and $x$, with $x\in H_{d-1}$.}
\label{fig10}
\end{center}
\end{figure}

In order to easily describe the geometry, it is convenient to focus first on two consecutive Poincar\'{e} patches in \textrm{AdS}$_{d+1}$ with $x^{-}>0$ and $x^{-}<0$, respectively. As shown in figure \ref{fig10}, these two regions are separated by the surface in \textrm{AdS}$_{d+1}$ defined by $x^{-}=0$ and are parameterized by the light cone coordinate $x^{+}$ and by a point $x$ in the transverse hyperbolic space $H_{d-1}$. Since the two Poincar\'{e} patches are invariant under translations generated by $x^{-}\partial_{\mu}+2x_{\mu}\partial_{+}$, we may parameterize the $x^{-}<0$ patch with new coordinates
\begin{align*}
& x^{+}-2\sigma\cdot x+\sigma^{2}x^{-}~,\\
& x^{-}~,\\
& x-\sigma x^{-}~,
\end{align*}
where $\sigma\in\mathbb{M}^{d}$ is arbitrary. We may then think of the two patches as being described by different coordinate systems glued along the
surface $x^{-}=0$ with the following gluing conditions
\begin{align}
x^{+}  & \rightarrow x^{+}+h\left(  x\right)  ~,\label{gluing}\\
x  & \rightarrow x~,\nonumber
\end{align}
where
\[
h\left(  x\right)  =2\sigma\cdot x~.
\]
As we move from one patch to the next, the light cone coordinate $x^{+}$ is shifted by an amount $h\left( x\right)$ which depends only on the transverse coordinates $x\in H_{d-1}$. Moreover, note that the function $h\left( x\right)$ satisfies
\begin{equation}
\left[  \square_{H_{d-1}}-\left(  d-1\right)  \right]
\,h=0~.\label{free}
\end{equation}
To show this fact, recall that any function $h\left( x\right)$ defined on a (pseudo) Euclidean space $\mathbb{E}^{D+2}$ which is harmonic and homogeneous, \textit{i.e.}, $\square_{\mathbb{E}^{D+2}}$ $h=0$ and $h\left( \lambda x\right) =\lambda^{-\alpha}h\left(  x\right)$, satisfies, when restricted to the (pseudo) sphere ${S}_{D+1}$ given by $x^{2}=\pm1$, the equation $\left[
\square_{{S}_{D+1}}-m^{2}\right]~h=0$, where $m^{2}=\pm\alpha\left( D-\alpha\right)$.

Up to now we have only described the original \textrm{AdS}$_{d+1}$ space using different coordinate systems in different parts of the space. The geometry describing a shock wave propagating along the surface $x^-=0$ is obtained by adding, to the vacuum Einstein equations, a source term localized at $x^-=0$ and independent of the null coordinate $x^+$. The shock geometry can then be easily described by gluing two \textrm{AdS}$_{d+1}$ patches as in (\ref{gluing}). As explained below, the gluing function now satisfies (\ref{free}) with a source term on the right--hand--side of the equation given by
\[
-16\pi G\,T\left(  x\right)\ ,
\]
where $x\in{H_{d-1}}$ and $G$ is the Newton constant measured in units of the \textrm{AdS} radius. We denote as usual with $\Pi\left( x,y\right)$ the Euclidean scalar propagator of mass $d-1$ in the transverse space $H_{d-1}$, canonically normalized so that
\[
\left[  \square_{H_{d-1}}-\left(  d-1\right)  \right]
~\Pi\left( x,y\right)  =-\delta\left(  x,y\right)  ~,
\]
and given explicitly in appendix A. In the presence of a source term $T\left( x\right)$ the gluing function $h\left( x\right)$ is then given by
\begin{equation}
h\left(  x\right)  =16\pi G\int_{H_{d-1}}\widetilde{dx^{\prime}}
~\Pi\left(  x,x^{\prime}\right)  T\left(  x^{\prime}\right)  ~.\label{hsol}
\end{equation}
The usual AdS Aichelburg--Sexl geometry can then be recovered by choosing a source due to a particle of energy $\omega$ localized in transverse space at
$y\in H_{d-1}$
\begin{align*}
T\left(  x\right)   & =\omega~\delta\left(  x,y\right)  ~,\\
h\left(  x\right)   & =16\pi G~\omega~\Pi\left(  x,y\right)  ~.
\end{align*}


\subsection{General Spin $j$ Interaction}


An equivalent way to present the shock wave geometry is to note that, as in flat space, the linear response of the metric to a stress--energy tensor localized along a null surface actually solves the full non--linear gravity equations. In this case, the full metric reads
\[
ds^{2}\left(  \mathrm{AdS}_{d+1}\right)  +\left(  dx^{-}\right)
^{2}\mathbf{h}\left(  \mathbf{x}\right)  ~,
\]
where $\mathbf{h}$ is localized on the shock front at $x^{-}=0$ and depends only on the transverse directions $x\in H_{d-1}$
\[
\mathbf{h}\left(  \mathbf{x}\right)  =\delta\left(  x^{-}\right)
h\left( x\right)  ~.
\]
The metric deformation $\mathbf{h}$ is generated by a stress--energy tensor
\begin{equation}
\mathbf{T}\left(  \mathbf{x}\right)  =\delta\left(  x^{-}\right)
T\left( x\right)  ~,\label{Tlightcone}
\end{equation}
located along the shock front. Einstein's equations
\[
\square_{\mathrm{AdS}_{d+1}}\mathbf{h}=-16\pi G~\mathbf{T}
\]
again translate, in transverse space, to
\[
\left[  \square_{H_{d-1}}-\left(  d-1\right)  \right]
~h=-16\pi G~T~,
\]
which is solved by (\ref{hsol}).

We now wish to consider the propagation of a complex scalar field $\Phi_{1}$ of mass $m_{1}$ in the presence of the shock. The metric deformation $\mathbf{h}$ changes the free Lagrangian
\[
\int\widetilde{d\mathbf{x}}~\Phi_{1}^{\star}\left(  \mathbf{\square}-m_{1}^{2}\right)  \Phi_{1}~
\]
by adding the minimal gravitational coupling $-4\int\widetilde{d\mathbf{x}}~\Phi_{1}^{\star}\mathbf{\partial}_{+}^{2}\Phi_{1}\mathbf{h}$, where we used that the only non--vanishing component of the metric fluctuations is $h_{--}=\mathbf{h}$. For the purposes of this paper, we will need to consider a more general interaction mediated by a spin $j$ particle. We may still consider a classical profile for $\mathbf{h}$ localized on the null surface $x^{-}=0$ as described above, but now associated to a shock wave of the spin $j$ massless field. The interaction with the scalar field $\Phi_{1}$ will then be of the more general form
\begin{equation}
-4\int\widetilde{d\mathbf{x}}~~\Phi_{1}^{\star}\mathbf{\partial}_{+}^{j}
\Phi_{1}\mathbf{~h~}, \label{coup1}
\end{equation}
where now $\mathbf{h}$ is the component $h_{-\cdots-}$ of the spin $j$ field. The equations of motion for $\Phi_{1}$ then read
\begin{equation}
\left(  \mathbf{\square}-m_{1}^{2}\right)  \Phi_{1}=4\mathbf{h}
~\mathbf{\partial}_{+}^{j}\Phi_{1}\mathbf{~}=4\delta\left(x^{-}\right)
h~\mathbf{\partial}_{+}^{j}\Phi_{1}~, \label{shockEQ}
\end{equation}
which translate in a boundary condition for $\Phi_{1}$ at the location of the shock. Around $x^-\sim 0$ the differential equation (\ref{shockEQ}) simplifies to
$$
\partial_+\partial_-\Phi_{1} =  - \delta\left(x^{-}\right)h~ \partial_+^j\Phi_{1}\ .
$$
Taking the Fourier transform $\int dx^+\, e^{-ix^+s}$ with respect to $x^+$, we obtain
$$
\partial_- \ln{\Phi}_1\left(s,x^-,x\right) =  - \delta\left(  x^{-}\right) (is)^{j-1} h(x)\ .
$$
Therefore, the value of the field $\Phi_{1}$ changes across the shock according to
\begin{equation}
\Phi_1\left(  x^{+},x\right)  \rightarrow e^{h\left(  x\right)
\partial_{+}^{j-1}}
\Phi_1\left(  x^{+},x\right)  ~.\label{glue}
\end{equation}
In particular, for $j=2$ we recover the previous result $x^{+}\rightarrow x^{+}+h\left(  x\right)  $ in (\ref{gluing}).


\section{Two--Point Function in the Shock Wave Geometry\label{BtoBSec}}


We are now in the position to compute the two--point function of the scalar field $\Phi_{1}$ in the presence of the shock. From now on we shall view the field $\Phi_{1}$ as dual to the operator $\mathcal{O}_{1}$ of conformal dimension $\Delta_{1}$, with $\Delta_{1}\left(  \Delta_{1}-d\right) =m_{1}^{\,2}$, and we shall be interested in its boundary to boundary correlator.

Recall first the standard bulk to boundary propagator $K_{\mathbf{p}_{1}}\left( \mathbf{x}\right)$ of $\Phi_{1}$, from a boundary point $\mathbf{p}_{1}$ to a bulk point $\mathbf{x}$ in the absence of the shock (\ref{gluing}). It is given by $\mathcal{C}_{\Delta_{1}}\left( -2\mathbf{x\cdot p}_{1}\right)^{-\Delta_{1}}$, where \footnote{The normalization $\mathcal{C}_{\Delta}$ is not the standard one used in the literature \cite{FreedmanRev, w98}. In this paper, the bulk to boundary propagator $K_{\mathbf{p}_{1}}\left( \mathbf{x}\right)$ is taken to be the limit of the bulk to bulk propagator $\mathbf{\Pi(x,y)}$ as the bulk point $\mathbf{y}$ approaches the boundary point $\mathbf{p}$. As shown in \cite{kw99, fmmr98} and briefly in appendix B, naive Feynman graphs in AdS computed with this prescription give correctly normalized CFT correlators, including the subtle two--point function.}
\[
\mathcal{C}_{\Delta}=\frac{1}{2\pi^{\frac{d}{2}}}\frac{\Gamma\left(
\Delta\right)  }{\Gamma\left(  \Delta-\frac{d}{2}+1\right)  }
\]
and where we must pay particular attention to the exact phase factor. More precisely, as described at the end of section 2, given the boundary point $\mathbf{p}_{1}$, the \textrm{AdS} space may be divided in an infinite sequence of Poincar\'{e} patches separated by the surfaces $-2\mathbf{x\cdot p}_{1}=0$ and labeled by integers $n$ increasing as we move forward in global time. The $n=0$ patch is the one which is spacelike related to the boundary point, as shown in figure \ref{DecompAdS}. Then, the correct definition of the bulk to boundary propagator is
\begin{equation}
K_{\mathbf{p}_{1}}\left(  \mathbf{x}\right)  =\mathcal{C}_{\Delta_{1}}
\frac{i^{-2\Delta_{1}\left\vert n\right\vert }}{\left\vert
2\mathbf{x\cdot p}_{1}\right\vert ^{\Delta_{1}}}~.\label{propGen}
\end{equation}
We shall mostly concentrate on the three patches labeled by $n=0,\pm1$ and shown in grey in figure \ref{DecompAdS}, where we can also write
\begin{equation}
K_{\mathbf{p}_{1}}\left(  \mathbf{x}\right)  =\frac{\mathcal{C}_{\Delta_{1}}
}{\left(  -2\mathbf{x\cdot p}_{1}+i\epsilon\right)  ^{\Delta_{1}}
}~.\label{ieps}
\end{equation}
Let us stress that (\ref{ieps}) is not valid in general. In fact, had we extended (\ref{ieps}) throughout the whole AdS space, we would have the propagator from a collection of boundary points related to $\mathbf{p}_{1}$ and $ -\mathbf{p}_{1}$ by $2\pi$ translations in global time.
\begin{figure}
[ptb]
\begin{center}
\includegraphics[height=1.7593in, width=2.3342in]
{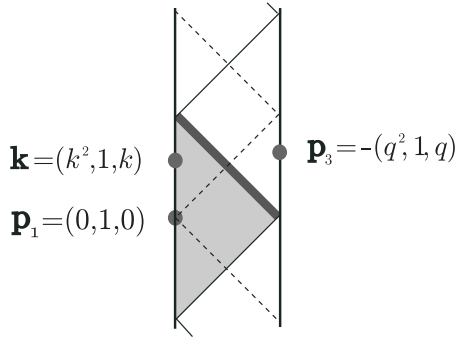}
\caption{Computation of the two--point function between the boundary points $\mathbf{p}_{1}$ and $\mathbf{p}_{3}$ in the presence of a shock along the thick diagonal line. This is equivalent to a linear superposition of propagators from
$\mathbf{k}$ to $\mathbf{p}_{3}$ in the absence of the shock, where the boundary point $\mathbf{k}$ runs over the grey patch $x^{-}>0$. We have divided the AdS space in patches along the $x^{-}=0$ surface (continuous lines, including the shock surface) and along the $x^{+}=0$ surface (dashed lines).}
\label{fig11}
\end{center}
\end{figure}

\begin{figure}
[ptb]
\begin{center}
\includegraphics[height=1.7593in]
{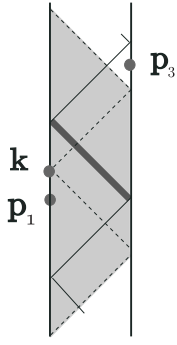}
\caption{In expression (4.6), we are not allowed to use (4.2) for the bulk to boundary propagator. In fact, the point $\mathbf{p}_3$, limiting boundary point of $\mathbf{x}$, is not always inside the $n=0,\pm 1$ patches of point $\mathbf{k}$, which are shown in grey. In general, $\mathbf{p}_3$ is within the $n=0,1,2$ patches of $\mathbf{k}$.}
\label{iepsilon}
\end{center}
\end{figure}

Now we compute the two--point function between boundary points $\mathbf{p}_{1}$ and $\mathbf{p}_{3}$ in the presence of a shock wave, where $\mathbf{p}_{1}$ is on the boundary of the patch preceding the shock, as shown in figure \ref{fig11}. Using translations $x^{-} \partial_{\mu}+2x_{\mu}\partial_{+}$ in this patch, we are free to place the point $\mathbf{p}_{1}$ at
\[
\mathbf{p}_{1}=\left(  0,1,0\right)  ~,
\]
so that the relevant bulk to boundary function is given, before the shock, by
\[
\frac{\mathcal{C}_{\Delta_{1}}}{\left(  x^{+}+i\epsilon\right)  ^{\Delta_{1}}
}~=\frac{i^{-\Delta_{1}}\mathcal{C}_{\Delta_{1}}}{\Gamma\left(
\Delta _{1}\right) }\int_{0}^{\infty}ds~s^{\Delta_{1}-1}e^{isx^{+}}~.
\]
Just after the shock, using the gluing relation (\ref{glue}), the scalar profile becomes
\begin{equation}
\frac{i^{-\Delta_{1}}\mathcal{C}_{\Delta_{1}}}{\Gamma\left(
\Delta _{1}\right) }\int_{0}^{\infty}ds~s^{\Delta_{1}-1}e^{isx^{+}+\left(
is\right)  ^{j-1}h\left(  x\right)  }~.\label{before}
\end{equation}
We want to write the above function, defined on the shock surface $x^{-}=0$,
as a coherent sum
\begin{equation}
\mathcal{C}_{\Delta_{1}}\int\frac{dk}{\left( 2\pi\right)^{d}}
~\frac{F\left(  k\right)  }{\left( x^{+}-2k\cdot
x+i\epsilon\right) ^{\Delta_{1}}} \label{after}
\end{equation}
of bulk to boundary propagators $K_{\mathbf{k}}\left( \mathbf{x}\right)$ where, as shown in figure \ref{fig11}, the point $\mathbf{k=}\left( k^{2},1,k\right)$ runs over the boundary of the patch preceding the shock. To determine $F\left( k\right)$ we equate the Fourier transforms with respect to $x^+$ of (\ref{before}) and (\ref{after}) and obtain, for $s>0$,
\[
e^{\left(  is\right)  ^{j-1}h\left(  x\right) }=\int\frac{dk}{\left( 2\pi\right)  ^{d}}~F\left(  k\right) ~e^{-2ik\cdot\left(  sx\right)  }~.
\]
This equation may now be inverted by considering $sx$ as a point in the future Milne wedge \textrm{M}, decomposed in its radial and angular parts. We then obtain
\begin{equation}
F\left(  k\right)  =2^{d}\int_{0}^{\infty}s^{d-1}ds\int_{H_{d-1}
}\widetilde{dx}~e^{2ik\cdot\left(  sx\right)  +\left(  is\right)
^{j-1}h\left(  x\right)  }~.\label{Tfun}
\end{equation}
Having obtained the scalar profile (\ref{after}) just after the shock, we may now evolve it forward after the surface $x^{-}=0$ and compute the boundary to boundary correlator $\mathcal{E}$, by considering the limit of the profile
\begin{equation}
\int\frac{dk}{\left(  2\pi\right)  ^{d}}~~F\left(  k\right)  ~K_{\mathbf{k}
}\left(  \mathbf{x}\right) \label{ss1}
\end{equation}
as the point $\mathbf{x}$ moves towards the boundary point
\[
\mathbf{p}_{3}=-\left(  q^{2},1,q\right)  ~
\]
in the patch after the shock. We need to be careful with phases using the general form of the propagator (\ref{propGen}) since $\mathbf{p}_{3}$ can be outside the $n=0,\pm 1$ Poincar\'e patches of $\mathbf{k}$, as shown in figure \ref{iepsilon}. More precisely, one arrives at the following integral
\[
\mathcal{E}=\mathcal{C}_{\Delta_{1}}\int\frac{dk}{\left(
2\pi\right)  ^{d}}~\frac{F\left(  k\right)  }{[-\left(  k-q\right)
^{2}+i\epsilon\operatorname{sign}\left(  k^{0}-q^{0}\right)
]^{\Delta_{1}}}~.
\]
Using (\ref{Tfun}) and changing variables to $k^{\prime}=s\left( k-q\right) $ we obtain
$$
\mathcal{E} = i^{-2\Delta_{1}} \mathcal{C}_{\Delta_{1}} \int_{0}^{\infty} s^{2\Delta_{1}-1} ds \int_{H_{d-1}} \widetilde{dx} ~ e^{2iq\cdot\left( sx \right) + \left( is \right)^{j-1} h \left(x\right) } \int \frac{dk^{\prime}}{\pi^{d}} ~ \frac{e^{2ik^{\prime} \cdot x}}{[k^{\prime2} - i\epsilon \operatorname{sign} \left( k^{\prime0} \right)]^{\Delta_{1}}}~.
$$
The last integral in the above expression is a constant since $x\in H_{d-1}$. This constant can be evaluated, for example, by considering the limit $h=0$ with $q\in-\mathrm{M} $ in the past Milne wedge. In this case $\mathcal{E}$ is given by the free propagator $\mathcal{C}_{\Delta_{1}}\left( -q^{2}\right) ^{-\Delta_{1}}$, thus showing that the constant is given by $\mathcal{N}_{\Delta_{1}}$. We have then arrived at the final result
\[
\mathcal{E}=i^{-2\Delta_{1}}\mathcal{C}_{\Delta_{1}}\mathcal{N}_{\Delta_{1}
}\int_{0}^{\infty}s^{2\Delta_{1}-1}ds\int_{H_{d-1}}\widetilde
{dx}~e^{2iq\cdot\left(  sx\right)  +\left(  is\right)
^{j-1}h\left( x\right)  }~.
\]
When $j=2$ the $s$ integral can be easily performed to obtain
\begin{equation}
\mathcal{E}=\mathcal{C}_{\Delta_{1}}\mathcal{N}_{\Delta_{1}}\Gamma\left(
2\Delta_{1}\right)
\int_{H_{d-1}}\frac{\widetilde{dx}~}{\left(
2q\cdot x+h\left(  x\right)  +i\epsilon\right)  ^{2\Delta_{1}}}\ ,
\qquad \left( j=2\right) . \label{sss3}
\end{equation}


\section{Creating the Shock Wave Geometry\label{CreateShock}}


In flat space the computation analogous to the previous section yields a non--perturbative approximation to the four--point amplitude in the eikonal kinematical regime. To understand the relation between the AdS result of the previous section and the dual CFT four--point function, we consider the tree level term in the expansion of the two--point function $\mathcal{E}$ in powers of $h$
\begin{align}
\mathcal{E}_{1} &=\left(  -\right)  ^{j-1}\mathcal{C}_{\Delta_{1}}
\mathcal{N}_{\Delta_{1}}\Gamma\left(  2\Delta_{1}+j-1\right)  \int
_{H_{d-1}}\widetilde{dx}~\frac{h\left(  x\right)
}{\left(  2q\cdot x+i\epsilon\right)  ^{2\Delta_{1}+j-1}}~
\nonumber
\\
\nonumber
\\
&
=16\pi G \left(  -\right)  ^{j-1}\mathcal{C}_{\Delta_{1}}
\mathcal{N}_{\Delta_{1}}\Gamma\left(  2\Delta_{1}+j-1\right)
\int_{H_{d-1}}\widetilde{dx}~\widetilde{dy}~
\frac{\Pi\left(x,y\right) T\left(y\right)} {\left(  2q\cdot
x+i\epsilon\right)  ^{2\Delta_{1}+j-1}}~. \label{TwoPTTree}
\end{align}
In this section, we shall show that the function $\mathcal{E}_{1}$ can also be computed from the Feynman graph in figure \ref{fig4}$(b)$, which corresponds to a tree level exchange of a massless spin $j$ particle between two scalar particles $\Phi_{1}$ and $\Phi_{2}$, with appropriate external wave functions. The fields $\Phi_{i}$ have masses $m_{i}$ and are dual to boundary operators $\mathcal{O}_{i}$ of conformal dimension $\Delta_{i}$. Moreover, the relevant coupling for the field $\Phi_{2}$, parallel to (\ref{coup1}), is given by \footnote{The sign $(-)^j$ indicates that, for odd $j$, the fields $\Phi_1$ and $\Phi_2$ are oppositely charged with respect to the spin $j$ interaction field. With this convention, graph \ref{fig4}$(b)$ corresponds to an attractive interaction, independently of $j$.}
\begin{equation}
-4\left( -\right)
^{j}\int\widetilde{d\mathbf{x}}~~\Phi_{2}^{\star
}\mathbf{\partial}_{-}^{j}\Phi_{2}\mathbf{~h}^{\prime}~,
\label{vertRel}
\end{equation}
where $\mathbf{h}'$ is the $h_{+\cdots+}$ component of the interaction spin $j$ field, which has a two--point function
\[
\left\langle \mathbf{h}\left(  \mathbf{x}\right)
\mathbf{h}^{\prime}\left( \mathbf{x}^{\prime}\right)
\right\rangle =8\pi G~\mathbf{\Pi}\left(
\mathbf{x,x}^{\prime}\right)  ~.
\]
The massless scalar propagator in \textrm{AdS}$_{d+1}$ is canonically normalized by $\square\mathbf{\Pi}\left( \mathbf{x,x}^{\prime}\right) =i\mathbf{\delta}\left(\mathbf{x,x}^{\prime}\right)$ and it is explicitly given in appendix A. We will denote with $\phi_{1},\phi_{3}$ and $\phi_{2},\phi_{4}$ the external wave functions of the AdS graph corresponding, respectively, to the two fields $\Phi_{1}$ and $\Phi_{2}$ interacting through the massless exchange. In general the external states $\phi_{1},\phi_{3}$ and $\phi_{2},\phi_{4}$ will be linear combinations of the bulk to boundary propagators
\begin{align*}
K_{\mathbf{p}}\left(  \mathbf{x}\right)   &
=\mathcal{C}_{\Delta_{1}}\left(
-2\mathbf{x\cdot p}\right)  ^{-\Delta_{1}}~,\\
\tilde{K}_{\mathbf{p}}\left(  \mathbf{x}\right)   & =\mathcal{C}_{\Delta_{2}
}\left(  -2\mathbf{x\cdot p}\right)  ^{-\Delta_{2}}~,
\end{align*}
respectively. Throughout this section, we will fix the external states $\phi_{1},\phi_{3}$ to be
\[
\phi_{1}=K_{\mathbf{p}_{1}}~,~\ \ \ \ \ \ \ \ \ \ \
~\phi_{3}=K_{\mathbf{p}_{3}}~,
\]
according to the previous section\footnote{In this section, all external states $\phi _{i}$ are built starting from the canonical bulk to boundary propagator (\ref{propGen}). This includes the states $\phi _{2}$, $\phi _{4}$ used to explicitly construct the shock wave. The phase conventions in (\ref{propGen}) show that the bulk to boundary propagator $K_{\mathbf{p}}\left( \mathbf{x}\right) $ corresponds to a non--normalizable wave with a $\delta $--function source at the boundary point $\mathbf{p}$. Computation of Feynman integrals with those external states corresponds then, in the dual CFT, to a computation of the
boundary correlator in the relevant Lorentzian regime. Note, finally, that the phases in (\ref{propGen}) are obtained by a canonical Wick rotation in global time, starting from Euclidean AdS in global coordinates.}. If we also choose
\[
\phi_{2}=\tilde{K}_{\mathbf{p}_{2}}~,~\ \ \ \ \ \ \ \ \ \ \ ~\phi_{4}
=\tilde{K}_{\mathbf{p}_{4}}~,
\]
the graph \ref{fig4}$(b)$ computes the amplitude $\mathcal{C}_{\Delta_{1}}\mathcal{C}_{\Delta_{2}}A_1\left( \mathbf{p}_{i}\right)=\mathcal{C}_{\Delta_{1}}\mathcal{C}_{\Delta_{2}}\mathbf{p}_{13}^{-\Delta_{1}}\mathbf{p}_{24}^{-\Delta_{2}}\mathcal{A}_{1}\left( \mathbf{p}_{i}\right)$. However, we shall choose the wave functions $\phi_{2},\phi_{4}$ so that the corresponding vertex (\ref{vertRel}), schematically given by $\partial_{\mu }^{j}\,\phi_{2}\phi_{4}$, is localized along the shock surface $x^{-}=0$ and only along the $\mu=-$ direction. In other words, the wave functions $\phi_2$ and $\phi_4$ will be chosen such that the vertex (\ref{vertRel}) is the source for the shock wave. This will be achieved by choosing $\phi_{2}$ and $\phi_{4}$ to be a particular linear combinations of the basic external wave functions $\tilde{K}_{\mathbf{p}}$. More precisely, the fields $\phi_{2}$ and $\phi_{4}$ will respectively vanish after and before the shock, so that their overlap $\partial_{\
 mu}^{j}~\phi_{2}\phi_{4}$ is supported only at $x^{-}=0$. Moreover, near $x^{-}\sim 0$, the functions $\phi_{2}$ and $\phi_{4}$ will be respectively chosen to behave in the light cone directions $x^{\pm}$ as $\left( x^{-}\right)^{\Delta_{2}+j-1}$ and $\left( x^{-}\right)^{-\Delta_{2}}$, so that their overlap $\partial_{-}^{j}\phi_{2}\phi_{4}$ goes as $1/x^{-}$ $\sim\delta\left( x^{-}\right)$. With this specific choice of external states $\phi_{2},\phi_{4}$, graph \ref{fig4}$(b)$ is explicitly given by
\[
-4i\int_{\mathrm{AdS}_{d+1}}\widetilde{d\mathbf{x}}~~\phi_{3}\partial_{+}^{j}\phi_{1}\mathbf{~h}(\mathbf{x})\, ,
\]
where
\begin{align}
\mathbf{h}\left(  \mathbf{x}\right)   & =16\pi G~i\int_{\mathrm{AdS}_{d+1}}\widetilde{d\mathbf{x}^{\prime}}~\ \mathbf{\Pi}\left(
\mathbf{x,x}^{\prime
}\right)  \mathbf{T}\left(  \mathbf{x}^{\prime}\right)  ~,\nonumber\\
\mathbf{T}  & =-2\left(  -\right)
^{j}\phi_{4}\partial_{-}^{j}\phi
_{2}~.\label{T24}
\end{align}
As discussed above, the functions $\phi_{2},\phi_{4}$ are chosen so that the source function $\mathbf{T}$ is supported on the light cone $\mathbf{T}\left( \mathbf{x}\right)  =\delta\left( x^{-}\right) T\left( x\right)$, and the graph \ref{fig4}$(b)$ computes $\mathcal{E}_{1}$ for the specific choice of transverse
source $T\left( x\right)$.
\begin{figure}
[ptb]
\begin{center}
\includegraphics[keepaspectratio,height=2in]
{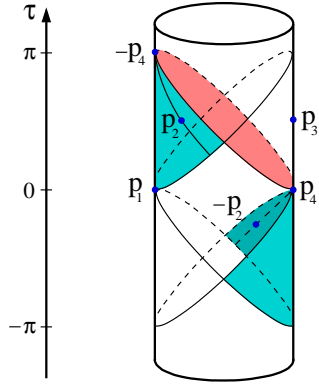} \caption{Construction of the external wave
functions $\phi_{2}$ and $\phi_{4}$ starting from the bulk to
boundary propagators $\tilde{K}_{\pm\mathbf{p}_{2}}$ and
$\tilde{K}_{\pm\mathbf{p}_{4}}$. The points $\pm\mathbf{p}_{4}$
are fixed, whereas the points $\pm\mathbf{p}_{2}$ are free to move
in the past Milne wedges (shaded regions of the boundary). In
particular, in (5.8), the points $\pm\mathbf{p}_{2}$ lie
along a ray from the origin, as shown. The source points
$\mathbf{p}_{1}$, $\mathbf{p}_{3}$ are as in figure 6.
We also show global AdS time.} \label{fig12}
\end{center}
\end{figure}

Following figure \ref{fig12}, we start by choosing as external
state $\phi_{4}$ the linear combination
\begin{align}
\phi_{4}  & =i^{2\Delta_{2}}~\tilde{K}_{-\mathbf{p}_{4}}-\tilde{K}
_{\mathbf{p}_{4}} \nonumber
\\
& =\mathcal{C}_{\Delta_{2}}\left[  (x^{-}-i\epsilon)^{-\Delta_{2}}
-(x^{-}+i\epsilon)^{-\Delta_{2}}\right]  ~, \label{phi4}
\end{align}
where we chose $\mathbf{p}_{4} = (1,0,0)$. The wave function
$\phi_4$ clearly vanishes before the shock for $x^{-}>0$.
Similarly, the function $\phi_{2}$ will be given by the general
linear combination
\begin{equation}
\phi_{2}=\int\frac{dp}{\left(  2\pi\right)  ^{d}}G\left(  p\right)
\left(
\tilde{K}_{\mathbf{p}_{2}}-i^{2\Delta_{2}}\tilde{K}_{-\mathbf{p}_{2}}\right)~,
\label{phi2}
\end{equation}
where we write $\mathbf{p}_{2}=(1,p^2,p)$. The integrand in
(\ref{phi2}) vanishes for $\mathbf{x\cdot p_2}<0$, so that, for
$G(p)$ supported in the past Milne wedge $-M$, the wave function
$\phi_2$ vanishes after the shock for $x^-<0$. Recall that we are
interested in the overlap $\phi_4\partial_-^j\phi_2$, so that we
need only the behavior of $\phi_2$ for $x^- \sim 0$. This in turn
is controlled only by $G(p)$ for $p\sim 0$. To show this, we shall
for a moment assume that $G(p)$ in (\ref{phi2}) is homogeneous in
$p$ as $G(\lambda p)=\lambda^c G(p)$. Then $\phi_2$ is an
eigenfunction of $x^{+}\partial_{+}-x^{-}\partial_{-}$ as follows
\begin{equation*}
\phi_{2}(  \lambda^{-1}x^{+},\lambda x^{-},x) =\lambda^{c-\Delta
_{2}+d}\phi_{2}(x^{+},x^{-},x)  ~,
\end{equation*}
and behaves, close to $x^-\sim 0$, as
$$
\phi_2(x^{+},x^{-},x)\simeq(x^-)^{c-\Delta_2+d}\phi_2(0,1,x)\, ,
$$
where $x\in H_{d-1}$. In general, we shall take, for reasons which
shall become clear shortly,
\begin{align*}
G\left( p\right) &= G_0 \left(
p\right) +\cdots\, ,\\
G_0\left( \lambda p\right) &= \lambda^{2\Delta_{2}+j-d-1} G_0
\left( p\right)\, ,
\end{align*}
where the dots denote sub--leading terms for $p\sim 0$. The above
discussion then immediately implies that the behavior of
$\phi_{2}$ just before the shock is given by
\[
\phi_{2}\left(  x^{+},x^{-},x\right)  \simeq\left(  x^{-}\right)
^{\Delta _{2}+j-1}g\left(x\right)+\cdots  ~,
\]
where $x$ is a position in transverse space $H_{d-1}$ and where
$g\left(x\right)$ is determined uniquely by $G_0(p)$.

We are now in a position to explicitly compute the source term
$\mathbf{T}$ in (\ref{T24}). We first recall the following
representation of the delta--function
\begin{align*}
& \Gamma\left(  \alpha\right)  \Gamma\left(  \beta\right)  [(x^{-}
-i\epsilon)^{-\alpha}-(x^{-}+i\epsilon)^{-\alpha}][(-x^{-}-i\epsilon)^{-\beta
}-(-x^{-}+i\epsilon)^{-\beta}]\\
& =\left\{
\begin{array}
[c]{c} -2\pi^{2}\delta\left(  x^{-}\right)  ~\ ,~\ \ \ \ \ \ \ \ \
\ \ \ \ \ \left(
\alpha+\beta=1\right)\ , \\
0~\ ,~\ \ \ \ \ \ \ \ \ \ \ \ \ \ \ \ \ \ \ \ \ \ \ \ \ \ \ \
\left( \alpha+\beta<1\right)\ .
\end{array}
\right.
\end{align*}
Writing the leading behavior of $\phi_{2}$ as
\[
\frac{\Gamma\left(  \Delta_{2}+j\right)  \Gamma\left(
1-\Delta_{2}-j\right) }{2\pi i} \Big[
(-x^{-}-i\epsilon)^{\Delta_{2}+j-1}-(-x^{-}+i\epsilon
)^{\Delta_{2}+j-1} \Big]  g\left(x\right)
\]
and using the above representation of $\delta\left( x^{-}\right)$
we conclude that the source function $\mathbf{T}$ in (\ref{T24})
is given by
\begin{align*}
\mathbf{T}\left(  \mathbf{x}\right)   & =\delta\left( x^{-}\right)
T\left(
x\right)  ~,\\
T\left(  x\right)   & =\left(  -\right)  ^{j-1}2\pi
i~\frac{\Gamma\left( \Delta_{2}+j\right)  }{\Gamma\left(
\Delta_{2}\right)  }g\left(x\right)  ~.
\end{align*}

To explicitly compute the function $g\left(x\right)$ in terms of
the weight function $G_0\left( p\right)$, we must simply evaluate
(\ref{phi2}) at $\left( 0,1,x\right)$, with $G$ replaced by $G_0$.
The first term in (\ref{phi2}) gives
\begin{align*}
& \mathcal{C}_{\Delta_{2}}\int\frac{dp}{\left(  2\pi\right)  ^{d}}
\frac{G_0\left(  p\right)  }{\left(  -1+2p\cdot x+i\epsilon\right)  ^{\Delta_{2}}}\\
& =\frac{\mathcal{C}_{\Delta_{2}}~i^{-\Delta_{2}}}{\Gamma\left(
\Delta _{2}\right)
}\int_{0}^{\infty}dt~t^{\Delta_{2}-1}~\int\frac{dp}{\left(
2\pi\right)  ^{d}}G_0\left(  p\right)  ~e^{it\left(  -1+2p\cdot x\right)  }\\
&
=\frac{\mathcal{C}_{\Delta_{2}}~i^{j-1}}{2^{2\Delta_{2}+j-1}}\frac
{\Gamma\left(  1-\Delta_{2}-j\right)  ~}{\Gamma\left(
\Delta_{2}\right) }G_0\left(  x\right)  ~,
\end{align*}
where we abuse notation and denote with $G_0\left( x\right)
=\left( 2\pi\right)^{-d}\int dp~e^{ip\cdot x}~G_0\left( p\right)$
the Fourier transform of $G_0\left(p\right)$. The second term in
(\ref{phi2}) is similarly given by
\[
i^{2\Delta_{2}}\mathcal{C}_{\Delta_{2}}\int\frac{dp}{\left(
2\pi\right) ^{d}}\frac{G_0\left(  p\right)  }{\left(  1-2p\cdot
x+i\epsilon\right)
^{\Delta_{2}}}=\frac{\mathcal{C}_{\Delta_{2}}~i^{1-j}}{2^{2\Delta_{2}+j-1}
}\frac{\Gamma\left(  1-\Delta_{2}-j\right)  ~}{\Gamma\left( \Delta
_{2}\right)  }G_0\left(  -x\right)  ~.
\]
Note that, in this case, the $i\epsilon$ prescription is correct
since $G_0\left(p\right)$ is supported only in the past Milne
wedge $-\mathrm{M}$. We finally conclude that the $T$--channel
exchange Witten diagram \ref{fig4}$(b)$ with external wave
functions $\phi_4$ and $\phi_2$, respectively as in (\ref{phi4})
and (\ref{phi2}), is given by (\ref{TwoPTTree}) with
\begin{equation}
T\left(  x\right)
=-\frac{2\pi~\mathcal{C}_{\Delta_{2}}^{2}}{2^{2\Delta
_{2}+j-1}}~\frac{\Gamma\left(  1-\Delta_{2}\right)  }{\Gamma\left(
\Delta _{2}\right)} \Big[  i^{j}G_0\left(  x\right)
+i^{-j}G_0\left( -x\right) \Big]  ~,\label{GtoT}
\end{equation}
where recall that we are interested in $x\in H_{d-1}$. Denoting
with $A_{1}^{\pm\pm}$ the tree level correlator associated to
graph \ref{fig4}$(b)$ when the external points are at
$\mathbf{p}_{1}$, $\mathbf{p}_{3}$ and $\pm\mathbf{p}_{2}$,
$\pm\mathbf{p}_{4}$, the same Witten diagram can be written as
\begin{equation}
\mathcal{E}_{1} =
\mathcal{C}_{\Delta_{1}}\mathcal{C}_{\Delta_{2}}\int\frac{dp}{\left(
2\pi\right)  ^{d}}G\left(  p\right)  \left(  i^{2\Delta_{2}}A_{1}
^{+-}+i^{2\Delta_{2}}A_{1}^{-+}-A_{1}^{++}-i^{4\Delta_{2}}A_{1}^{--}\right)~.
\label{Int4pt}
\end{equation}

It is particularly convenient to choose a weight function $G\left(
p\right)$ supported along a straight line as shown in figure \ref{fig12}
\begin{equation}
G\left(  p\right)  =\int_0^a dt~\ t^{2\Delta_{2}+j-2}~\left(
2\pi\right) ^{d}\delta^{d}\left(  p-\hat{p}t\right)
~,\label{Gline}
\end{equation}
with $\hat{p}\in-H_{d-1}$ a unit vector. Note that the behavior of
$G(p)$ for $p\sim 0$ is independent of $a$, and the leading
behavior $G_0(p)$ is obtained by setting $a=\infty$ in
(\ref{Gline}). We then have
\[
G_0\left(  x\right)  =i^{2\Delta_{2}+j-1}\frac{\Gamma\left(
2\Delta _{2}+j-1\right)  }{\left(  \hat{p}\cdot x+i\epsilon\right)
^{2\Delta_{2} +j-1}}~,
\]
and finally, for $x\in H_{d-1}$,
\[
T\left(  x\right)  =\left(  -\right)
^{j-1}\pi~\mathcal{C}_{\Delta_{2}
}\mathcal{N}_{\Delta_{2}}~\Gamma\left(  2\Delta_{2}+j-1\right)
\frac {1}{\left(  2\hat{p}\cdot x\right)  ^{2\Delta_{2}+j-1}}~.
\]
For this particular source the two--point function
(\ref{TwoPTTree}) becomes
\begin{align}
\mathcal{E}_{1} &= 16\pi^2 G
\mathcal{C}_{\Delta_{1}}\mathcal{C}_{\Delta_{2}}
\mathcal{N}_{\Delta_{1}}\mathcal{N}_{\Delta_{2}} \Gamma\left(
2\Delta_{1}+j-1\right) \Gamma\left(  2\Delta_{2}+j-1\right)\times
\nonumber
\\
&\times\int_{H_{d-1}}\widetilde{dx}~\widetilde{dy}~
\frac{\Pi\left(x,y\right)} {\left(  2q\cdot x\right)
^{2\Delta_{1}+j-1} \left(  2\hat{p}\cdot y\right)
^{2\Delta_{2}+j-1}}\ , \label{partE1}
\end{align}
where we recall that both $q\cdot x$ and $\hat{p}\cdot y$ are
positive if $q$ is in the past Milne wedge $-\mathrm{M}$.


\section{Relation to the Dual CFT Four--Point Function\label{secsix}}


In this section we shall express the Lorentzian four--point
correlators in (\ref{Int4pt}) in terms of the Euclidean
four--point function by means of analytic continuation. We will
denote with $\mathcal{A}_{1}^{\pm\pm}\left( z,\bar{z}\right)$ the
Lorentzian amplitudes corresponding to the tree level correlators
$A_{1}^{\pm\pm}$. More precisely, we have
\[
A_{1}^{\pm\pm}=\frac{i^{-2\Delta_{1}\left\vert n_{13}\right\vert }
i^{-2\Delta_{2}\left\vert n_{24}^{\pm\pm}\right\vert }}{\left\vert
\mathbf{p}_{13}\right\vert ^{\Delta_{1}}\left\vert
\mathbf{p}_{24}\right\vert
^{\Delta_{2}}}\mathcal{A}_{1}^{\pm\pm}~.
\]
We have been careful with the exact phases and introduced, as in
the discussion of the bulk to boundary propagator in section
(\ref{BtoBSec}), the integer numbers $n_{ij}$ which label the
Poincar\'e patch, relative to point $\mathbf{p}_{i}$, containing
the point $\mathbf{p}_{j}$. Note that the cross--ratios
$z$,$\bar{z}$ are invariant under re--scalings
$\mathbf{p}_{i}\rightarrow\lambda_{i}\mathbf{p}_{i}$ with
$\lambda_{i}$ arbitrary, and in particular they are independent of
the choice of signs in $\pm\mathbf{p}_{2}$, $\pm\mathbf{p}_{4}$.
This means that the functions $\mathcal{A}_{1}^{\pm\pm }\left(
z,\bar{z}\right)$ are given by specific analytic continuations of
the basic Euclidean four--point amplitude $\mathcal{A}_{1}\left(
z,\bar{z}\right)$.

Without loss of generality we may fix from now on
$q\in-\mathrm{M}$. Recalling that $G\left( p\right)$ is
non--vanishing only for $p\in-\mathrm{M}$, we have that
$n_{13}=0$, $n_{24}^{++}=0$, $n_{24}^{--}=2$,
$n_{24}^{+-}=n_{24}^{-+}=1$. Therefore we can write (\ref{Int4pt})
as
\[
\mathcal{E}_{1}=\mathcal{C}_{\Delta_{1}}\mathcal{C}_{\Delta_{2}}\int\frac
{dp}{\left(  2\pi\right)  ^{d}}\frac{G\left(  p\right)  }{\left(
-q^{2}\right)  ^{\Delta_{1}}\left(  -p^{2}\right)
^{\Delta_{2}}}\left(
\mathcal{A}_{1}^{+-}+\mathcal{A}_{1}^{-+}-\mathcal{A}_{1}^{++}-\mathcal{A}
_{1}^{--}\right)  ~,
\]
where we recall that $z,\bar{z}$ are implicitly defined by
\[
z\bar{z}=q^{2}p^{2}~,~\ \ \ \ \ \ \ \ \ \ z+\bar{z}=2q\cdot p~.
\]
In particular, choosing $q=\hat{q}\in-H_{d-1}$ of unit norm and
$G\left( p\right)$ as in (\ref{Gline}), we obtain the expression
\begin{equation}
\mathcal{E}_{1}=\mathcal{C}_{\Delta_{1}}\mathcal{C}_{\Delta_{2}}\int_{0}
^{a}dt~\ t^{~j-2}\left(  \mathcal{A}_{1}^{+-}+\mathcal{A}_{1}^{-+}
-\mathcal{A}_{1}^{++}-\mathcal{A}_{1}^{--}\right)  ~,\label{cutEQ}
\end{equation}
where now
\begin{align}
z &  =-tw^{-1/2},~~\ \ \ \ \ \ \bar{z}=-tw^{1/2}~,\label{ss2}\\
w^{1/2}+w^{-1/2} &  =-2\hat{q}\cdot\hat{p}~.\nonumber
\end{align}
\begin{figure}
[ptb]
\begin{center}
\includegraphics[height=1.951in]
{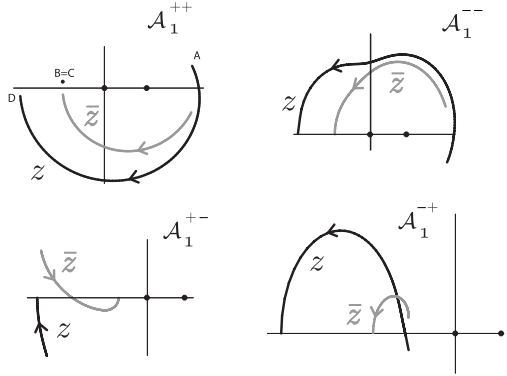} \caption{Wick rotation of $z$,$\bar{z}$ when the
external points are at $\mathbf{p}_{1}$, $\mathbf{p}_{3}$,
$\pm\mathbf{p}_{2}$, $\pm\mathbf{p}_{4}$. On the Euclidean
principal sheet of the amplitude we have $\bar{z}=z^{\star}$
(initial points of the curves), while after Wick rotating to the
Lorentzian domain $z,\bar{z}$ are real and negative (final
points). Points $A$, $B$, $C$ and $D$ refer to the detailed
analysis of $\mathcal{A}_{1}^{++}$ in figure 10.}
\label{fig1}
\end{center}
\end{figure}
Notice that for $\hat{q},\hat{p}\in-H_{d-1}$ we have that $-\hat
{q}\cdot\hat{p}\geq1$ and therefore both $z$ and $\bar{z}$ are
real and negative. We now must consider more carefully the issue of the analytic
continuation. Consider a generic boundary point $\mathbf{p}$, and
let $\tau$ be the decompactified global time. We have that
\begin{align*}
p^{d+1} &  =\lambda  \cos\left(  \tau\right)  ~,\\
p^{0} &  =\lambda \sin\left(  \tau\right)  .
\end{align*}
In particular, denoting with $\tau_{1}$, $\tau_{3}$,
$\tau_{2}^{\pm}$, $\tau_{4}^{\pm}$ the global times of the four
boundary points $\mathbf{p}_{1}$, $\mathbf{p}_{3}$,
$\pm\mathbf{p}_{2}$, $\pm\mathbf{p}_{4}$, we clearly have that
(see figure \ref{fig12})
\begin{align*}
\tau_{1}  & =\tau_{4}^{+}=0,\\
\tau_{4}^{-}  & =\pi,\\
0  & \leq\tau_{2}^{+},\tau_{3}\leq\pi~,\\
-\pi & \leq\tau_{2}^{-}\leq0~.
\end{align*}
We can then consider, for each of the boundary points under
consideration, the standard Wick rotation $\tau\rightarrow-i\tau$
parameterized by $0\leq \theta\leq1$
\begin{align*}
p^{d+1} &  =\lambda  \cos\left(  -i\tau e^{\frac{i\pi}
{2}\theta}\right)  ~,\\
p^{0} &  =\lambda \sin\left(  -i\tau e^{\frac{i\pi}
{2}\theta}\right)  ~,
\end{align*}
where $\theta=0$ corresponds to the Euclidean regime and
$\theta=1$ to the Minkowski setting. In particular, given the four
points $\mathbf{p}_{1}$, $\mathbf{p}_{3}$, $\pm\mathbf{p}_{2}$,
$\pm\mathbf{p}_{4}$, we may follow the cross--ratios $z\left(
\theta\right)$, $\bar{z}\left( \theta\right)$ as a function of
$\theta$. The plots of $z\left( \theta\right)$, $\bar{z}\left(
\theta\right)$ in the four cases $\mathcal{A}_{1}^{\pm\pm}$ are
shown in figure \ref{fig1}. Note that, in the Euclidean limit, we
have that
\[
\bar{z}\left(  0\right)  =\left(  z\left(  0\right)  \right)
^{\star}~,
\]
as expected. On the other hand, when $\theta=1$, the cross--ratios
$z\left( 1\right), \bar{z}\left( 1\right)$ are explicitly given by
(\ref{ss2}). We remark that, although figure \ref{fig1} has been
derived with a specific choice of $p=t\hat{p}$ and $q=\hat{q}$,
the qualitative features of the curves are independent of the
chosen $p$, $q$. Moreover, from figure \ref{fig1}, we deduce that
\begin{align*}
\mathcal{A}_{1}^{++}\left(  z,\bar{z}\right)   &  =\mathcal{A}_{1}
^{\circlearrowright}\left(  z-i\epsilon,\bar{z}-i\epsilon\right)  ~,\\
\mathcal{A}_{1}^{+-}\left(  z,\bar{z}\right)   &
=\mathcal{A}_{1}\left(
z-i\epsilon,\bar{z}-i\epsilon\right)  ~,\\
\mathcal{A}_{1}^{-+}\left(  z,\bar{z}\right)   &
=\mathcal{A}_{1}\left(
z+i\epsilon,\bar{z}+i\epsilon\right)  ~,\\
\mathcal{A}_{1}^{--}\left(  z,\bar{z}\right)   &  =\mathcal{A}_{1}
^{\circlearrowleft}\left(  z-i\epsilon,\bar{z}+i\epsilon\right) ~,
\end{align*}
where $\mathcal{A}_{1}^{\circlearrowright}$
($\mathcal{A}_{1}^{\circlearrowleft}$) is the analytic
continuation of the Euclidean amplitude $\mathcal{A}_{1}$ obtained
by keeping $\bar{z}$ fixed and by transporting $z$ clockwise
(counterclockwise) around the point at infinity. We explain the
above result by concentrating on the first case of
$\mathcal{A}_{1}^{++}$. Consider the black curve $z\left(
\theta\right)$ in figure \ref{fig1} for $\mathcal{A}_{1}^{++}$. It
can be deformed without crossing any singularity to the black
curve in figure \ref{fig13}, which is composed of three parts AB,
BC and CD. The first part AB is just the complex conjugate of the
curve $\bar{z}\left( \theta\right)$. Therefore, at point B, we are
clearly on the principal sheet $\mathcal{A}_{1}$. The curve BC, on
the other hand, rotates clockwise around the point at $\infty$,
moving therefore to the sheet
$\mathcal{A}_{1}^{\circlearrowright}$. Finally the last segment CD
is immaterial, since the function $\mathcal{A}_{1}$ is only singular at $z=0,1,\bar{z}$.
\begin{figure}
[ptb]
\begin{center}
\includegraphics[width=1.7in]
{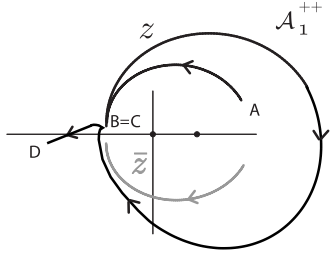} \caption{The black curve going through points A,B,C,D
is obtained from a continuous deformation of the curve $z(\theta)$
in figure 9 for $\mathcal{A}_1^{++}$. It shows the
relation of $\mathcal{A}_1^{++}$ with the Euclidean amplitude
$\mathcal{A}_1$.} \label{fig13}
\end{center}
\end{figure}

In general we have that the Euclidean amplitude is real, in the
sense that
\[
\mathcal{A}_{1}\left( z,\bar{z}\right)  =\mathcal{A}_{1}\left(
\bar {z},z\right)  ~,~\ \ \ \ \ \ \ \ \ \ \
\mathcal{A}_{1}^{\star}\left( z,\bar{z}\right)
=\mathcal{A}_{1}\left( z^{\star},\bar{z}^{\star}\right)  ~.
\]
Therefore
\[
\mathcal{A}_{1}^{\circlearrowleft}\left( z,\bar{z}\right) =\left[
\mathcal{A}_{1}^{\circlearrowright}\left(
z^{\star},\bar{z}^{\star}\right) \right]  ^{\star}~.
\]
We then conclude that
\[
\mathcal{A}_{1}^{+-}+\mathcal{A}_{1}^{-+}-\mathcal{A}_{1}^{++}-\mathcal{A}
_{1}^{--}=4\pi\operatorname{Im}\mathcal{M}_{1}\left(
z-i\epsilon,\bar {z}-i\epsilon\right)  ~,
\]
where $\mathcal{M}_{1}$ is the discontinuity function of $\mathcal{A}_{1}$ defined in the introduction.
Thus, the right hand side of (\ref{cutEQ}) is explicitly given
by
\begin{equation}
4\pi\mathcal{C}_{\Delta_{1}}\mathcal{C}_{\Delta_{2}}\int_{0+i\epsilon}
^{a}dt~\ t^{~j-2}~\operatorname{Im}\mathcal{M}_{1}\left( -tw^{1/2}
,-tw^{-1/2}\right)  ~. \label{cut2}
\end{equation}
Recall from the discussion in the previous section that the above
integral is independent of $a$. Therefore, the integrand is
supported at $t=0$, and the leading behavior of $\mathcal{M}_{1}$
is given by
\begin{equation}
\mathcal{M}_{1}\left(  z,\bar{z}\right)  \simeq\left(  -z\right)
^{1-j} M \left( \frac{\bar{z}}{z} \right)~,~ \label{itislate}
\end{equation}
with $M^{\star}(w)=M(w^{\star})$. Note, in particular, that the
residue function $M\left(  w\right)  $ must be real in order for
the integrand in (\ref{cut2}) to be localized at $t=0$, which
follows from the independence of the integral on the upper limit
of integration $a$. Then (\ref{cut2}) becomes
\[
4\pi\mathcal{C}_{\Delta_{1}}\mathcal{C}_{\Delta_{2}}w^{\frac{j-1}{2}}M\left(
w\right)  \int_{0}^{a}dt~\operatorname{Im}\frac{1}{t+i\epsilon}
~=-2\pi^{2}\mathcal{C}_{\Delta_{1}}\mathcal{C}_{\Delta_{2}}w^{\frac{j-1}{2}
}M(w)~.
\]
The two--point function $\mathcal{E}_{1}$ is, on the other hand,
given by (\ref{partE1}) with $q=\hat{q}\in-H_{d-1}$. This gives
then an integral representation for $M\left( w\right)$ 
\begin{align}
w^{\frac{j-1}{2}}M(w) &
=-8G\mathcal{N}_{\Delta_{1}}\mathcal{N}_{\Delta_{2} }\Gamma\left(
2\Delta_{1}-1+j\right)  \Gamma\left( 2\Delta_{2}-1+j\right)
\times\nonumber\\
&
\times~\int_{H_{d-1}}\widetilde{dx}\widetilde{dy}~\frac{\Pi\left(
x,y\right)  }{\left(  2\hat{q}\cdot x\right)
^{2\Delta_{1}-1+j}\left( 2\hat{p}\cdot y\right)
^{2\Delta_{2}-1+j}}~.\label{result}
\end{align}
where we recall that $w^{1/2}+w^{-1/2}=-2\hat{q}\cdot\hat{p}$.
Clearly we have that
\[
M(w) = w^{1-j} M \left(\frac{1}{w}\right).
\]
Finally, using (\ref{itislate}) and (\ref{result}),  we obtain the
equivalent result (\ref{tt11}) for the leading behavior of
$\mathcal{M}_{1}$ given in the introduction.


\subsection{An Example in $d=2$\label{SectExample}}


Let us conclude this section with a simple example where we can check our result. We shall consider the case $j=0$ corresponding to massless scalar exchange in $\mathrm{AdS}$. The basic amplitude $\mathcal{A}_{1}$ is given by
\[
\frac{\mathcal{C}_{\Delta_{1}}\mathcal{C}_{\Delta_{2}}}{\mathbf{p}
_{13}^{\Delta_{1}}\mathbf{p}_{24}^{\Delta_{2}}}\mathcal{A}_{1}=-4i\int
_{\mathrm{AdS}_{d+1}}\widetilde{d\mathbf{x}}\frac{\mathcal{C}_{\Delta_{1}}
^{\,2}}{\left(  -2\mathbf{x\cdot p}_{1}\right)  ^{\Delta_{1}}\left(
-2\mathbf{x\cdot p}_{3}\right)  ^{\Delta_{1}}}~\mathbf{h}\left(
\mathbf{x}\right)  ~,
\]
where
\[
\mathbf{h}\left(  \mathbf{x}\right)  =-32i\pi G~\int_{\mathrm{AdS}_{d+1}
}\widetilde{d\mathbf{y}}~\mathbf{\Pi}\left(  \mathbf{x,y}\right) \
\frac{\mathcal{C}_{\Delta_{2}}^{\,2}}{\left(  -2\mathbf{y\cdot
p}_{2}\right)
^{\Delta_{2}}\left(  -2\mathbf{y\cdot p}_{4}\right)  ^{\Delta_{2}}}
\]
and where $\mathbf{\Pi}\left( \mathbf{x,y}\right)$ is the massless propagator in $\mathrm{AdS}_{d+1}$. We shall concentrate, in particular, on the simple case $d=\Delta_{2}=2$, so that the scalar field dual to the operator $\mathcal{O}_{2}$ is massless in $\mathrm{AdS}_{3}$. In this case we can use the general technique in \cite{Rastelli2} and easily compute
\[
\mathbf{h}\left( \mathbf{x}\right) = -\frac{1}{\left( 2\pi\right)^{2}
}\frac{8\pi G}{~\mathbf{p}_{24}\left( -2\mathbf{x\cdot
p}_{2}\right)  \left( -2\mathbf{x}\cdot\mathbf{p}_{4}\right)  }~,
\]
where we have used $\mathcal{C}_{\Delta}=1/\left( 2\pi\right)$ for $d=2$. In terms of the standard functions
\[
D_{\Delta_{i}}^{d}\left(  \mathbf{p}_{i}\right)  =\int_{H_{d+1}
}\mathbf{~}~\frac{\mathbf{~\widetilde{d\mathbf{x}}}}{
{\textstyle\prod\nolimits_{i}}
\left(  -2\mathbf{x\cdot p}_{i}\right)  ^{\Delta_{i}}}~,~
\]
reviewed in appendix A, we conclude, after Wick rotation
\[
\int_{\mathrm{AdS}_{d+1}}\widetilde{d\mathbf{x}}\rightarrow-i\int
_{H_{d+1}}\widetilde{d\mathbf{x}}~,
\]
that
\begin{equation}
\mathcal{A}_{1}=\frac{8G}{\pi}~\mathbf{p}_{13}^{\Delta_{1}}\mathbf{p}
_{24}~D_{\Delta_{1},\Delta_{1},1,1}^{2}\left(  \mathbf{p}_{1},\mathbf{p}
_{3},\mathbf{p}_{2},\mathbf{p}_{4}\right)  ~. \label{ampex}
\end{equation}

We may also explicitly compute the integral (\ref{tt11}), which controls the leading behavior as $z,\bar{z}\rightarrow0$
of $\mathcal{M}_{1}=\operatorname*{Disc}_{z} \mathcal{A}_{1}$. For $\Delta_{2}=2$, $j=0$ we can use again the methods of \cite{Rastelli2} to explicitly perform the $y$--integral in (\ref{tt11}). We then arrive at the result
\[
\mathcal{M}_{1}\simeq-8G~\frac{\Gamma\left(  2\Delta_{1}-1\right)  }
{\Gamma\left(  \Delta_{1}\right)  ^{2}}\left(  -q^{2}\right)  ^{\Delta_{1}
}\left(  -p^{2}\right)  ~D_{2\Delta_{1}-1,1}^{0}\left(
-q,-p\right)  ~.
\]
Using the explicit form of $D_{2\Delta_{1}-1,1}^{0}$ given in appendix A, we conclude that $\mathcal{M}_{1}\simeq-zM\left( \bar{z}/z\right)$, where the function $M\left( w\right)$ is explicitly given by
\begin{equation}
M\left(  w\right)  =-\frac{4G}{2\Delta_{1}-1}\,w F \left(  1,\Delta_{1}
,2\Delta_{1} \Big| 1-w\right) , \label{sss1}
\end{equation}
where $F$ is the standard hypergeometric function.

Furthermore, we shall restrict our attention to the special case $\Delta_{1}=2$, where the amplitude (\ref{ampex}) can be explicitly computed \cite{Bianchi}
\[
\mathcal{A}_{1}\left(  z,\bar{z}\right)
=-8G\frac{z^{2}\bar{z}^{2}}{\left(
\bar{z}-z\right)  }\left[  \frac{1}{1-\bar{z}}\partial-\frac{1}{1-z}
\bar{\partial}\right]  a\left(  z,\bar{z}\right)  ~,
\]
with
\[
a\left(  z,\bar{z}\right)  =\frac{\left(  1-z\right)  \left(
1-\bar {z}\right)  }{\left(  z-\bar{z}\right)  }\left[
\mathrm{Li}_{2}\left( z\right)  -\mathrm{Li}_{2}\left(
\bar{z}\right)  +\frac{1}{2}\mathrm{\ln }\left(  z\bar{z}\right)
\mathrm{\ln}\left(  \frac{1-z}{1-\bar{z}}\right) \right]  ~
\]
and where $\mathrm{Li}_{2}\left(  z\right)  $ is the dilogarithm. It is easy to check that $a\left( z,\bar{z}\right) = a\left( z^{-1},\bar{z}^{-1}\right)$, so that we quickly deduce that
\[
\operatorname*{Disc}{}_{z}\,a\left( z,\bar{z}\right) =-\frac{1}{2}
\frac{\left( 1-z\right) \left( 1-\bar{z}\right)  }{\left(
z-\bar
{z}\right) }\;\mathrm{\ln}\left( \frac{1-z}{1-\bar{z}}\cdot\frac{\bar{z}}
{z}\right)  ~.
\]
Applying to $\operatorname*{Disc}{}_{z}\,a$ the differential operator relating $a$ with $\mathcal{A}_{1}$, we obtain an exact expression for the discontinuity of $\mathcal{A}_{1}$
\[
\mathcal{M}_{1}=4G\frac{z\bar{z}}{\left( z-\bar{z}\right)
^{3}}\left[
z^{2}-\bar{z}^{2}+\ln\left( \frac{1-z}{1-\bar{z}}\cdot\frac{\bar{z}}
{z}\right) \left( 2z\bar{z}-z\bar{z}^{2}-z^{2}\bar{z}\right)
\right]  ~.
\]
For small $z,\bar{z}$ the above expression simplifies to $\mathcal{M}_{1}\simeq-zM\left( \bar{z}/z\right)$, with
\[
M\left( w\right) =-4G\frac{w}{\left( 1-w\right)^{3}}\Big[ 1+2w\ln\left( w\right) -w^{2}\Big]  ~.
\]
This is exactly (\ref{sss1}) for $\Delta_{1}=2$.


\section{Shock Wave in the BTZ Black Hole\label{BTZsec}}


In this final section, we will extend the previous analysis of the
two--point function in the presence of a shock to the case of the
Schwarzschild BTZ black hole. Therefore, throughout this section,
we shall work in $d=2$. The main interest of this calculation is
to understand, within the BTZ black hole example, how spacelike
singularities and horizons can be described in terms of CFT
amplitudes. This problem was first addressed in \cite{m01}, where
the BTZ geometry is conjectured to be dual to an entangled state
between two copies of the CFT located at the two disconnected BTZ
boundaries, and later further explored in \cite{kos02, fhks03,
fl05, br03, hlr06}. In particular, the main goal of \cite{kos02}
is to extract, from CFT correlators, information on the physics
behind the horizon, with particular emphasis on the singularity.
One can probe physics behind the horizon by studying the
two--point function $\left\langle
{\mathcal{O}}_{1}(\mathbf{p}_{1}){\mathcal{O}}_{1}(\mathbf{p}_{3})\right\rangle$ of a boundary operator, where the boundary points are located on the two distinct BTZ asymptotic boundaries. In the case where this operator creates a bulk scalar particle with a large mass $m_{1}$ (and therefore large conformal dimension) one may evaluate the two--point function in the semiclassical geodesic approximation as \cite{kos02}
\[
\left\langle
{\mathcal{O}}_{1}(\mathbf{p}_{1}){\mathcal{O}}_{1}(\mathbf{p}
_{3})\right\rangle \simeq\exp\left(  -m_{1}{\mathcal{L}}\right)~,
\]
where ${\mathcal{L}}$ is the (regularized) proper length of the
spacelike geodesic which connects the two boundary points. Such a
correlator gives access to the full spacetime, including the
region behind the horizon. Extensions of these ideas in $d=5$ were
carried out in \cite{fhks03, fl05}. On a different line
\cite{moz98}, these two--point functions may also be used in the
computation of greybody factors for BTZ black holes.

In what follows, we shall extend the results of \cite{kos02} and
compute the two--point function in the presence of a shock wave
along the black hole horizon. As for the pure AdS case, this
should be related to a specific kinematical regime, dominated by
gravitational exchange, of the full four--point function in the
entangled thermal state of the CFT. Therefore, this computation
contains non--perturbative information about the dual CFT, which
is probed, beyond the semiclassical gravitational regime, at
finite $G$. These results could then allow us to study, following
reasonings along the lines of \cite{kos02, fhks03, fl05}, the
physics of the singularity in a full quantum gravity regime. We
shall leave a full investigation of these issues to future work,
limiting ourselves to the computation of the shock two--point
function.

As described in section 2, Anti--de Sitter space AdS$_{3}$ is
given by the $\mathbb{M}^{2}\times\mathbb{M}^{2}$ embedding
\[
-\mathbf{x}^{2}=x^{+}x^{-}+x\bar{x}=1~,
\]
where we use light cone coordinates $x,\bar{x}$ on the second
$\mathbb{M}^{2}$. As is well know \cite{BTZ, BTZ22}, the
Schwarzschild BTZ black hole is described by the identifications
\begin{equation}
x\sim e^{\alpha}x~,\ \ \ \ \ \ \ \bar{x}\sim
e^{-\alpha}\bar{x}~,\label{cici1}
\end{equation}
where $\alpha$ is related to the black hole mass $M$ by
$\alpha=2\pi\sqrt{M}$. The region outside the horizon, with
$x^{-}>0$ and $x^{+}<0$, can be parameterized with coordinates
$r,t,\theta$ as
\begin{align}
x^{\pm}  & =\mp e^{\pm t}\sqrt{r^{2}-1}~,\nonumber\\
x  & =re^{\theta},\label{cici2}\\
\bar{x}  & =re^{-\theta}~.\nonumber
\end{align}
with metric
\[
ds^{2}=-\left(  r^{2}-1\right)
dt^{2}+\frac{dr^{2}}{r^{2}-1}+r^{2}d\theta ^{2}~.
\]
The BTZ identification (\ref{cici1}) simply amounts to the
periodicity $\theta\sim\theta+\alpha$.

The identifications (\ref{cici1}) clearly leave the surface
$x^{-}=0$ invariant, and therefore we may still construct a shock
geometry along the horizon by considering the gluing condition
\[
x^{+}\rightarrow x^{+}+h\left( \theta\right)  ~,
\]
where, at the shock $x^{-}=0$, $r=1$, we have
\thinspace$x=e^{\theta}$, $\bar{x}=e^{-\theta}$. Clearly, in order
to preserve (\ref{cici1}), we must have that
\[
h\left( \theta+\alpha\right) =h\left( \theta\right)  ~.
\]
If we let $\omega$ be the energy of the particle creating the
shock and located at $\theta=0$, the function $h\left(
\theta\right)$ satisfies
\[
\left( \frac{\partial^{2}}{\partial\theta^{2}}-1\right)
h(\theta)=-16\pi G\omega \;{\sum_{n}}\, \delta\left(
\theta-\alpha n\right)
\]
with $n\in\mathbb{Z}$, whose particular solution, satisfying the
periodicity condition, is given by \cite{Sfetsos}
\[
h(\theta)=8\pi G\omega \;{\sum_{n}} e^{-|\theta-\alpha n|}~.
\]

In order to compute the two--point function across the shock, we
must extend the results of section \ref{ShockSec}. Indeed, in
section \ref{BtoBSec} we have used the invariance under
translations in the Poincar\'{e} patch $x^{-}>0$ to place the
source point $\mathbf{p}_{1}$ at the origin $\left( 0,1,0\right)$
of $\mathbb{M}^{2}$. This is clearly not general enough in the
present context, since we are considering the quotient of
$\mathbb{M}^{2}$  by a boost. We must therefore consider the more
general source point
\[
\mathbf{p}_{1}=\left( -e^{t},e^{-t},e^{\theta},e^{-\theta}\right)
~,
\]
where the last two entries explicitly denote the light cone
coordinates on $\mathbb{M}^{2}$. Moreover, we must consider a
source which is invariant under (\ref{cici1}), so we must add all
points $\mathbf{p}_{1}$ with $\theta
\rightarrow\theta+\alpha\mathbb{Z}$. We also define the probe
point $\mathbf{p}_{3}$ after the shock by
\[
\mathbf{p}_{3}=\left(
e^{t^{\prime}},-e^{-t^{\prime}},e^{\theta^{\prime}
},e^{-\theta^{\prime}}\right)  ~,
\]
where we parameterize the region with $x^{-}<0$ and $x^{+}>0$
after the shock using (\ref{cici2}) with the only change
$x^{\pm}=\pm e^{\pm t}\sqrt{r^{2}-1}$. We then have that
\[
-2\mathbf{p}_{1}\cdot\mathbf{p}_{3}=2\left(  \cosh\left(
t-t^{\prime}\right) +\cosh\left(  \theta-\theta^{\prime}\right)
\right)  ~.
\]
Therefore, in the absence of a shock, the basic two--point
function of a field of conformal dimension $\Delta_{1}$ is given
by \cite{kos02}
\begin{equation}
\left\langle \mathcal{O}_{1}(  t,\theta)
\mathcal{O}_{1}(t^{\prime},\theta^{\prime})  \right\rangle
=\mathcal{C}_{\Delta_{1}} \sum_{n}\frac{1}{\Big(  2\cosh\left(
t-t^{\prime}\right) +2\cosh\left(
\theta+n\alpha-\theta^{\prime}\right)  \Big)
^{\Delta_{1}}}~.\label{sss4}
\end{equation}
On the other hand, when the gluing function $h\left(
\theta\right)$ is non--vanishing, we may use (\ref{sss3}) to
obtain directly the two--point function. More precisely, to obtain
the vector $q$ in (\ref{sss3}) we first re--scale
$\mathbf{p}_{1}\rightarrow e^{t}\mathbf{p}_{1}$ and
$\mathbf{p}_{3}\rightarrow e^{t^{\prime}}\mathbf{p}_{3}$ in order
to rescale the $p^{-}$ coordinates to $\pm1$. Then the vector $q$
is the $\mathbb{M}^{2}$ part of
$-e^{t^{\prime}}\mathbf{p}_{3}-e^{t}\mathbf{p}_{1}$, which has
light cone components
$-(e^{t^{\prime}+\theta^{\prime}}+e^{t+\theta},e^{t^{\prime}-\theta^{\prime}}+e^{t-\theta})$.
Summing over images of the initial source point, we then obtain
the final result
\begin{align*}
\left\langle \mathcal{O}_{1}(  t,\theta) \mathcal{O}_{1}(
t^{\prime},\theta^{\prime}) \right\rangle _{\mathrm{shock}}  &
=\mathcal{C}_{\Delta_{1}}\mathcal{N}_{\Delta_{1}}\Gamma\left(
2\Delta
_{1}\right)  ~e^{\Delta_{1}\left(  t+t^{\prime}\right)  }~\times\\
& \times\sum_{n}\int\frac{d\chi}{\Big(  2e^{t}\cosh\left(
\theta+\alpha n-\chi\right)  +2e^{t^{\prime}}\cosh\left(
\theta^{\prime}-\chi\right) +h\left(  \chi\right)  \Big)
^{2\Delta_{1}}}~,
\end{align*}

\noindent which extends (\ref{sss4}) to the full BTZ shock
geometry. We shall leave a full exploration of the above result,
including its relation to the full four--point function in the
entangled thermal state of the CFT, for future research.


\section{Future Work \label{Conj}}


This paper is a  first step towards the understanding of the eikonal approximation in the context of the AdS/CFT correspondence. We were able to understand the AdS eikonal kinematical regime at tree level, relating the two--point function in the presence of a shock wave to the \textit{discontinuity} across a kinematical branch cut of the dual CFT four--point function associated to the Witten diagram
in figure \ref{fig4}$(b)$. Thus, in order to fully reconstruct the four--point amplitude, extra information is needed: the monodromy alone is not enough. As such, the understanding of the full eikonal re--summation is still missing and we leave it for future study. Nonetheless, in a companion paper \cite{Paper2} we explore the CFT consequences of the main result here obtained, and conjecture a possible resolution to the present problem concerning the reconstruction of the full four--point function.

Let us then conclude with other future directions of investigation:

\begin{itemize}

\item All the discussion of this paper has been done for pure \textrm{AdS}. It is possible that the full discussion can be generalized to \textrm{AdS} $\times$ \textrm{compact spaces} and to superspaces. This is important for applications of our results in the specific realizations of the AdS/CFT correspondence.

\item We have considered purely field theoretic interactions, neglecting all string effects. In flat space, on the other hand, it is known that, in the eikonal limit, the leading string effects simply Reggeize the gravitational interaction lowering its effective spin $j$ from $2$ to $2+\alpha^{\prime}t$. Reggeized interactions can then be re--summed as usual. It would be important to include this leading correction in the context of \textrm{AdS} physics, following the work of \cite{PolStra, PolStra22}.

\item Eikonal formul\ae \ have an even greater range of validity in flat space, as shown in \cite{ACV}. In the presence of a full string theoretic formulation of the interactions, the phase shift becomes an operator defining an explicitly unitary eikonal $S$--matrix. String effects are more than just Reggeization, but are still under control. The extension of these results to \textrm{AdS} would then be the next logical step, after the previous points have been understood.

\item It is of fundamental importance to extend the results of this paper to the BTZ geometry. In particular, one should be able to relate the two--point function computed in section \ref{BTZsec} to a four--point function in the BTZ background. If this program can be carried to completion, it would yield information about thermal correlators at finite $G$, which should probe spacetime, and in particular the singularity, in a truly quantum gravity regime.

\item Finally, it would be of outmost importance to test all these results against computations performed directly in the CFT duals at finite $N$, possibly at weak coupling.

\end{itemize}

\section*{Acknowledgments}
Our research is supported in part by INFN, by the MIUR--COFIN contract 2003--023852, by the EU contracts MRTN--CT--2004--503369, MRTN--CT--2004--512194, by the INTAS contract 03--51--6346, by the NATO grant PST.CLG.978785 and by the FCT contract POCTI/FNU/38004/2001. LC is supported by the MIUR contract \textquotedblleft Rientro dei cervelli\textquotedblright \ part VII. MSC was partially supported by the FCT grant SFRH/BSAB/530/2005. JP is funded by the FCT fellowship SFRH/BD/9248/2002. \emph{Centro de F\'{\i}sica do Porto} is partially funded by FCT through the POCTI program.

\vfill

\eject

\appendix


\section{Propagators and Contact $n$--Point Functions}


Let $\mathbf{x}$, $\mathbf{y}$ be two points in $\mathrm{AdS}_{d+1}$ or $H_{d+1}$, satisfying $\mathbf{x}^{2}=\mathbf{y}^{2}=-1$. Define the chordal distance
\[
z=-\frac{1}{4}\left(  \mathbf{x-y}\right)^{2}=\frac{1}{2}\left( 1+\mathbf{x\cdot y}\right)  ~.
\]
The scalar propagator $\mathbf{\Pi}\left( \mathbf{x,y}\right)$ of a scalar field of mass $m$, normalized to
\begin{align*}
\left[  \square_{\mathrm{AdS}_{d+1}}-m^{2}\right]
\,\mathbf{\Pi}\left( \mathbf{x,y}\right)   &
=i~\mathbf{\delta}\left(  \mathbf{x,y}\right)
~,~\ \ \ \ \ \ \ \ \ \ \ \ \ \ \ \ \ \ \text{(on }\mathrm{AdS}_{d+1}\text{)}\ ,\\
\left[  \square_{H_{d+1}}-m^{2}\right]
\,\mathbf{\Pi}\left( \mathbf{x,y}\right)   &
=-~\mathbf{\delta}\left(  \mathbf{x,y}\right)
~,~~\ \ \ \ \ \ \ \ \ \ \ \ \ \ \ \ \text{(on }H_{d+1}\text{)}\ ,
\end{align*}
is given explicitly by
$$
\mathbf{\Pi} = \frac{1}{\left(4\pi\right)^{\frac{d+1}{2}}} \frac {\Gamma \left(  \Delta \right) \Gamma \left( \frac{2\Delta-d+1}{2} \right)}{\Gamma \left( 2 \Delta-d+1 \right)}\ \frac{1}{\left(-z\right) ^{\Delta}}\ F \left. \left( \Delta, \frac{2\Delta-d+1}{2}, 2\Delta-d+1 \right| \frac{1}{z}\right)~,~
$$
where $\Delta$ is the conformal dimension of the dual boundary operator $2\Delta=d+\sqrt{d^{2}+4m^{2}}$. In particular, in this paper, we shall mostly denote with $\mathbf{\Pi}\left( \mathbf{x,y}\right)$ the Minkowskian massless propagator in $\mathrm{AdS}_{d+1}$ and with $\Pi\left( \mathbf{x,y}\right)$ the Euclidean massive propagator on $H_{d-1}$ of conformal dimension $d-1$. They are explicitly given by
$$
\mathbf{\Pi} = \frac{\Gamma \left( \frac{d+1}{2} \right)}{d\left(4\pi\right)^{\frac{d+1}{2}}} \ \frac{1}{\left(-z\right) ^{d}}\ F \left. \left( d, \frac{d+1}{2}, d+1 \right| \frac {1}{z}\right)
$$
and by
$$
\Pi = \frac{ \Gamma \left( \frac{d+1}{2}
\right)}{d(d-1)\left(4\pi\right)^{\frac{d-1}{2}}}\
\frac{1}{\left(-z\right)^{d-1}}\ F \left. \left( d-1,
\frac{d+1}{2}, d+1 \right| \frac{1}{z} \right).
$$

We also introduce the contact $n$--point functions
\[
D_{\Delta_{i}}^{d}\left(  \mathbf{p}_{i}\right)  =\int_{H_{d+1}
}\mathbf{~\widetilde{d\mathbf{y}}~\ }
{\textstyle\prod\nolimits_{i}}
~\frac{1}{\left(  -2\mathbf{y\cdot p}_{i}\right)  ^{\Delta_{i}}}~.
\]
They are given by the integral representation
\begin{align*}
D_{\Delta_{i}}^{d}\left(  \mathbf{p}_{i}\right)   & =\pi^{\frac{d}{2}}
\frac{\Gamma\left(  \frac{\Delta-d}{2}\right)  }{
{\textstyle\prod\nolimits_{i}}
\Gamma\left(  \Delta_{i}\right)  }\int\mathbf{~}
{\textstyle\prod\nolimits_{i}}
dt_{i}~\ t_{i}^{\Delta_{i}-1}~e^{\mathbf{-}\frac{1}{2}\sum_{i,j}t_{i}
t_{j}~\mathbf{p}_{ij}}\\
& =\frac{\pi^{\frac{d}{2}}}{2}\frac{\Gamma\left(
\frac{\Delta-d}{2}\right)
\Gamma\left(  \frac{\Delta}{2}\right)  }{
{\textstyle\prod\nolimits_{i}}
\Gamma\left(  \Delta_{i}\right)  }\int\mathbf{~}
{\textstyle\prod\nolimits_{i}}
dt_{i}~t_{i}^{\Delta_{i}-1}\ \frac{\delta\left(
{\textstyle\sum\nolimits_{i}}
t_{i}-1\right)  ~}{\left(  \frac{1}{2}\sum_{i,j}t_{i}t_{j}~\mathbf{p}
_{ij}\right)  ^{\frac{\Delta}{2}}}~,
\end{align*}
with $\Delta={\textstyle\sum\nolimits_{i}}\Delta_{i}$ and with $\mathbf{p}_{ij}=-2\mathbf{p}_{i}\cdot\mathbf{p}_{j}$. The two--point function integral can be explicitly computed as
\[
D_{\Delta_{1},\Delta_{2}}^{d} \left( \mathbf{p}_{1}, \mathbf{p}_{2} \right) =\frac{1}{\left( -\mathbf{p}_{1}^{2}\right)^{\frac{\Delta_{1}}{2}} \left( -\mathbf{p}_{2}^{2}\right)^{\frac{\Delta_{2}}{2}}}\ \frac{\pi^{\frac{d}{2}}}{2}\ \frac{\Gamma\left( \frac{\Delta-d}{2}\right) \Gamma\left( \frac{\Delta }{2}\right)}{\Gamma\left( \Delta\right)}\ \alpha^{\Delta_{2}}\ F \left. \left( \Delta_{2},\frac{\Delta}{2},\Delta \right| 1-\alpha^{2}\right)  ~,
\]
where
\[
\alpha+\frac{1}{\alpha}=\frac{-2\mathbf{p}_{1}\cdot\mathbf{p}_{2}}{\left(
-\mathbf{p}_{1}^{2}\right)  ^{\frac{1}{2}}\left(
-\mathbf{p}_{2}^{2}\right) ^{\frac{1}{2}}}~.
\]
Moreover, as shown in \cite{Bianchi}, the four--point function $D_{2,2,1,1}^{2}$ can be explicitly computed in terms of standard one loop box integrals, as reviewed in section \ref{SectExample}.


\section{Explicit Computations in Poincar\'{e} Coordinates}


The reader who is familiar with the calculation of correlation
functions in AdS in, \textit{e.g.}, \cite{FreedmanRev, w98,
fmmr98}, will find in the present appendix a pedagogical and very
explicit introduction to the calculations of the shock two--point
function performed in the bulk of the paper.

Let us recall two standard sets of coordinates in
$\mathrm{AdS}_{d+1}$. We first have the usual Poincar\'{e}
coordinates $\left( z,z^{\mu }\right) $, parameterizing the
Poincar\'{e} patch $x^{-}>0$. They are defined by
\[
\mathbf{x}=\left( x^{+},x^{-},x^{\mu }\right) =\frac{1}{z}\left(
z^{2}+z^{\mu }z_{\mu },1,z^{\mu }\right) ~,
\]
with the AdS metric taking the standard conformally flat form
\[
ds^{2}\left( \mathrm{AdS}_{d+1}\right) =\frac{dz^{2}+dz^{\mu }dz_{\mu }}{
z^{2}}~.
\]
Another useful set of coordinates, parameterizing the region
$x^{\mu }x_{\mu }=x^{+}x^{-}-1<0$ near the shock, is given by null
coordinates $\left( u^{+},u^{-},u^{\mu }\right) $ with $u^{\mu
}\in H_{d-1}$, where now

\begin{eqnarray}
x^{+} &=&\frac{4u^{+}}{4+u^{+}u^{-}}~,~\ \ \ \ \ \ \ \ \ \ \ \ \ x^{-}=\frac{
4u^{-}}{4+u^{+}u^{-}}~,  \label{suca2} \\
x^{\mu } &=&\frac{4-u^{+}u^{-}}{4+u^{+}u^{-}}~u^{\mu }~.
\nonumber
\end{eqnarray}

\noindent Here, $u^{\mu }$ denotes a point along the transverse
hyperboloid, which is parameterized in general by $d-1$ angular
variables. The AdS metric now reads
\[
ds^{2}\left( \mathrm{AdS}_{d+1}\right)
=-\frac{16du^{+}du^{-}}{\left( 4+u^{+}u^{-}\right) ^{2}}+\left(
\frac{4-u^{+}u^{-}}{4+u^{+}u^{-}}\right) ^{2}ds^{2}\left(
H_{d-1}\right) ~,
\]
with volume form given by
\begin{equation}
\epsilon _{\mathrm{AdS}_{d+1}}=8\frac{\left( 4-u^{+}u^{-}\right) ^{d-1}}{
\left( 4+u^{+}u^{-}\right) ^{d+1}}du^{+}~du^{-}~\epsilon _{H_{d-1}}~.  \label{vform}
\end{equation}
We shall mostly work in $d=2$, where
\[
u^{0}\pm u^{1}=e^{\pm \chi },~\ \ \ \ \ \ \ \ \ \ ds^{2}\left(H_{1}\right) =d\chi ^{2}~,~\ \ \ \ \ \ \ \ \epsilon_{H_{1}}=d\chi ~.
\]

We now recall the standard computation of the AdS
boundary--to--boundary two--point function, starting with
Poincar\'{e} coordinates. We parameterize points on the boundary
as always with $p_{1}^{\mu }$, where the boundary point is given
in global embedding coordinates by
\[
\mathbf{p}_{1}=\left( p_{1}^{\mu }p_{1\mu },1,p_{1}^{\mu }\right)
~.
\]
Since
\[
-2\mathbf{p}_{1}\mathbf{\cdot x=}\frac{1}{z}\left( z^{2}+\left(
z^{\mu }-p_{1}^{\mu }\right) ^{2}\right)
\]
we obtain the usual expression for the bulk--to--boundary
propagator
\[
K\left( z,z^{\mu };p_{1}^{\mu }\right) =\mathcal{C}_{\Delta }\left( \frac{z}{
z^{2}+\left( z^{\mu }-p_{1}^{\mu }\right) ^{2}}\right) ^{\Delta }.
\]
Given boundary data $\phi _{0}(p_{1}^{\mu })$, the bulk scalar
field value is given in terms of the propagator by \cite{w98,
fmmr98}
\[
\phi (z,z^{\mu })=\int d^{d}p_{1}\ K\left( z,z^{\mu };p_{1}^{\mu
}\right) \phi _{0}(p_{1}^{\mu }).
\]
Throughout the paper, the propagator $K$ is taken to be the limit
of the bulk--to--bulk propagator $\mathbf{\Pi }$, and its
normalization differs from the standard one in the literature
\cite{FreedmanRev, w98} by a factor of $2\Delta -d$. Therefore
$\lim_{z\rightarrow 0}z^{\Delta -d}\phi (z,z^{\mu })=\left(
2\Delta -d\right) ^{-1}\phi _{0}(z^{\mu })$. Moreover, the
two--point function is given by
\[
\left\langle {\mathcal{O}}(p_{1}^{\mu }){\mathcal{O}}(p_{2}^{\mu
})\right\rangle =\frac{\left( 2\Delta -d\right) }{\Delta
}\lim_{\epsilon \rightarrow 0}\int d^{d}z\ \epsilon ^{1-d}K\left(
\epsilon ,z^{\mu };p_{1}^{\mu }\right) \left. \frac{\partial
}{\partial z}K\left( z,z^{\mu };p_{2}^{\mu }\right) \right\vert
_{z=\epsilon }~,
\]
and it follows that

\begin{equation}
\left\langle {\mathcal{O}}(p_{1}){\mathcal{O}}(p_{2})\right\rangle =\frac{
\mathcal{C}_{\Delta }}{\left\vert p_{1}-p_{2}\right\vert ^{2\Delta
}}. \label{suca1}
\end{equation}

\noindent Recall that \cite{fmmr98} the coefficient $\left(
2\Delta -d\right) /\Delta $ arises from a careful treatment of
regularization at the boundary.

In the following we shall compute the two--point function, in the
presence of the shock wave, using null coordinates. It is then
instructive to repeat
the calculation above in these coordinates. A point on the boundary with $
u^{+}u^{-}=-4$ is given by
\[
\mathbf{p}_{1}=\left(
-\frac{2}{u_{1}^{-}},\frac{u_{1}^{-}}{2},u_{1}^{\mu }\right) ,
\]
and the bulk--to--boundary propagator is now given by
\[
\frac{\mathcal{C}_{\Delta }}{\left( -2\mathbf{p\cdot x}\right) ^{\Delta }}=
\mathcal{C}_{\Delta }\left[ \frac{\left( 4+u^{+}u^{-}\right) u_{1}^{-}u^{-}}{
2u^{+}u^{-}\left( u_{1}^{-}\right) ^{2}-8~\left( u^{-}\right)
^{2}-2\left( 4-u^{+}u^{-}\right) u_{1}^{-}u^{-}~u_{1}\cdot
u}\right] ^{\Delta }~.
\]%
Sending the point $\mathbf{x}$ to $\mathbf{p}_{2}$ on the boundary
one obtains the two--point function in null coordinates as
\[
\left\langle {\mathcal{O}}(u_{1}^{-},u_{1}^{\mu }){\mathcal{O}}
(u_{2}^{-},u_{2}^{\mu })\right\rangle =\mathcal{C}_{\Delta }\left[ \frac{
u_{1}^{-}u_{2}^{-}}{\left( u_{1}^{-}u_{1}^{\mu
}-u_{2}^{-}u_{2}^{\mu }\right) ^{2}}\right] ^{\Delta }~.
\]

In the presence of a shock the original metric in null coordinates
gains an additional term
\[
ds^{2}\left( \mathrm{AdS}_{3}\right) +\left( du^{-}\right)
^{2}\delta \left( u^{-}\right) h\left( \chi \right) ~,
\]
where from now on we specialize to the case $d=2$ for
concreteness. Recall that the function $h\left( \chi \right) $
satisfies
\[
\left( \frac{\partial ^{2}}{\partial \chi ^{2}}-1\right) h(\chi
)=-16\pi G\omega \delta \left( \chi \right) ~,
\]
so that the particular solution satisfying $\lim_{\chi \rightarrow
\pm \infty }h(\chi )=0$ is given by
\[
h(\chi )=8\pi G\omega e^{-|\chi |}.
\]
We now have all the required data in order to proceed with the
calculation of the two--point function in the AdS shock wave
background. The geometry is described by a metric $g_{mn}+\delta
g_{mn}$, where only $\delta g_{--}$ is non--vanishing and where
$g^{--}=0$. Therefore, the volume form is insensitive to $\delta
g_{mn}$ and is given by $\epsilon _{\mathrm{AdS}_{3}}$ in
(\ref{vform}). One can also compute the inverse metric, $\left(
g+\delta
g\right) ^{mn}=g^{mn}-\delta g^{mn}$, with single non--vanishing component $
~\delta g^{++}=4\delta \left( u^{-}\right) h\left( \chi \right) $.
The action for a scalar field in the AdS shock wave background is
then
\[
S[\phi ]=S_{0}[\phi ]+\frac{1}{2}\int \epsilon
_{\mathrm{AdS}_{3}}\left( \delta g^{mn}\partial _{m}\phi \partial
_{n}\phi \right)
\]
where $S_{0}[\phi ]$ is the standard AdS action for a scalar
field, and the new term has support on the shock wave alone, as it
includes a delta--function restricting it to $u^{-}=0$. This new
term is what we shall call the\textit{\ shock wave vertex}; it is
a new 2--vertex needed to compute the two--point function in the
AdS shock wave background, and it is precisely located at the
position of the shock wave. It is given explicitly by
\[
\int du^{+}d\chi \ h\left( \chi \right) ~\partial _{+}\phi
\partial _{+}\phi ~.~
\]
One may now compute the leading correction to the two--point function (\ref
{suca1}) in the AdS shock wave background. At a graphical level
this is rather simple: the leading\ graph includes two
boundary--to--bulk propagators, which meet at the shock wave
2--vertex; the higher order\ graphs includes two
boundary--to--bulk propagators, and $n$ different shock wave
2--vertices connected by $n-1$ bulk--to--bulk propagators. The
leading correction is simply given by
\[
2\mathcal{C}_{\Delta }^{2}\int du^{+}d\chi \ h\left( \chi \right)
~\partial _{+}\left( -2\mathbf{p}_{1}\mathbf{\cdot x}+i\epsilon
\right) ^{-\Delta }\partial _{+}\left(
-2\mathbf{p}_{2}\mathbf{\cdot x}+i\epsilon \right) ^{-\Delta }~,
\]
where $\mathbf{x}$ is given explicitly in terms of $u^{\pm },u^{\mu }$ in (
\ref{suca2}) and where the boundary points
$\mathbf{p}_{1}\mathbf{,p}_{2}$ are explicitly given by
$\mathbf{p}_{1}=\left( p_{1}^{\mu }p_{1\mu },1,p_{\mu }\right) $
and $\mathbf{p}_{2}=-\left( p_{2}^{\mu }p_{2\mu },1,p_{2}^{\mu
}\right) $. Working always in null coordinates, and using that,
along the shock at $u^{-}=0$ we have
\begin{eqnarray*}
\left( -2\mathbf{p}_{1}\mathbf{\cdot x}+i\epsilon \right) ^{-\Delta } &=&
\frac{i^{-\Delta }}{\Gamma \left( \Delta \right) }\int
ds_{1}~s_{1}^{\Delta
-1}e^{is_{1}\left( u^{+}-2u\cdot p_{1}\right) }~, \\
\left( -2\mathbf{p}_{2}\mathbf{\cdot x}+i\epsilon \right) ^{-\Delta } &=&
\frac{i^{-\Delta }}{\Gamma \left( \Delta \right) }\int
ds_{2}~s_{2}^{\Delta -1}e^{is_{2}\left( -u^{+}+2u\cdot
p_{2}\right) }~,
\end{eqnarray*}
we obtain
\[
2\mathcal{C}_{\Delta }^{2}\frac{i^{-2\Delta }}{\Gamma \left(
\Delta \right) ^{2}}\int ds_{1}ds_{2}~s_{1}^{\Delta }s_{2}^{\Delta
}\int du^{+}d\chi \ h\left( \chi \right) e^{i\left(
s_{1}-s_{2}\right) u^{+}}e^{2i\left( s_{2}p_{2}-s_{1}p_{1}\right)
\cdot u}~.
\]
Integrating in $u^{+}$ we obtain $2\pi \delta \left(
s_{1}-s_{2}\right) $
and therefore we finally get
\begin{eqnarray*}
&&4\pi \mathcal{C}_{\Delta }^{2}\frac{i^{-2\Delta }}{\Gamma \left(
\Delta \right) ^{2}}\int ds~s^{2\Delta }\int d\chi \ h\left( \chi
\right)
e^{2is\left( p_{2}-p_{1}\right) \cdot u}~ \\
&=&-2\pi \mathcal{C}_{\Delta }\mathcal{N}_{\Delta }\Gamma \left(
2\Delta +1\right) \int d\chi \ \frac{h\left( \chi \right) }{\left[
2\left( p_{2}-p_{1}\right) \cdot u\right] ^{2\Delta +1}}~~,
\end{eqnarray*}
which matches exactly the result (\ref{TwoPTTree}) in the bulk of
the paper. In this particular two--dimensional case one may
further compute explicitly the $\chi $ integral as
\[
8\pi G\omega \int d\chi \ \frac{e^{-|\chi |}~}{\left[ -qe^{-\chi }-\bar{q}
e^{\chi }\right] ^{2\Delta +1}}~,
\]
where
\[
q,\bar{q}=\left( p_{2}^{0}-p_{1}^{0}\right) \pm \left(
p_{2}^{1}-p_{1}^{1}\right) ~.
\]
Computing the integral one obtains the result(here $F$ is the standard $
{}_{2}F_{1}$ hypergeometric function),

\begin{eqnarray*}
-G\omega \frac{8\pi ^{2}\mathcal{C}_{\Delta }\mathcal{N}_{\Delta
}\Gamma \left( 2\Delta +1\right) }{\left( \Delta +1\right)}&&
\left[ \frac{1}{\left(
-q\right) ^{2\Delta +1}}\ F\left( \Delta +1,2\Delta +1,\Delta +2\left\vert -
\frac{\bar{q}}{q}\right) \right. +\right.  \\
&&\left. +\frac{1}{\left( -\bar{q}\right) ^{2\Delta +1}}\ F\left(
\Delta +1,2\Delta +1,\Delta +2\left\vert -\frac{q}{\bar{q}}\right)
\right. \right] ~.
\end{eqnarray*}

In order to proceed to higher orders in $G$, obtaining the exact
two--point function in the AdS shock wave background, we would
need to compute graphs with an arbitrary number of shock wave
vertices. On the other hand, this immediately poses a problem in
the calculation. A general graph includes bulk to bulk propagators
between the vertices which are positioned along the shock surface.
Since the geodesic distance of two points along the shock is
insensitive to the $v$ coordinate, the naive computation of the
graph produces divergences coming from the integrations along the
light cone coordinate $v$. This means that, in order to compute
higher order contributions to the shock wave two--point function,
one first needs to devise a suitable regularization of these
graphs. In section \ref{BtoBSec} we have solved this problem using
a generalization of the technique introduced by 't Hooft in
\cite{tHooft}.


\vfill

\eject

\bibliographystyle{plain}

\end{document}